\documentclass[%
 reprint,
superscriptaddress,
bibnotes,
 amsmath,amssymb,
 aps,
pra,
]{revtex4-2}

\usepackage{graphicx}
\usepackage{dcolumn}
\usepackage{bm}
\usepackage{physics}
\usepackage{orcidlink}
\usepackage{amsmath}
\usepackage{bbm}
\usepackage{xcolor}
\usepackage[most]{tcolorbox}
\usepackage[T1]{fontenc}
\usepackage{hyperref}

\begin{document}

\preprint{APS/123-QED}

\title{Nonlinear Optics Mediated by Chiral Waveguide QED: Generation of Momentum-anticorrelated Photon Pairs}

\author{Wenqi Tong}
 \email{tong76@purdue.edu}
 \affiliation{Elmore Family School of Electrical and Computer Engineering, Purdue University, West Lafayette, Indiana 47907, USA}
 
\author{Spencer J Walsh}
  \affiliation{Department of Physics and Astronomy, Purdue University, West Lafayette, Indiana 47907, USA}

\author{H. Alaeian}%
 \email{halaeian@purdue.edu}
 \affiliation{Elmore Family School of Electrical and Computer Engineering, Purdue University, West Lafayette, Indiana 47907, USA}
 \affiliation{Department of Physics and Astronomy, Purdue University, West Lafayette, Indiana 47907, USA}
 \affiliation{Purdue Quantum Science and Engineering Institute, Purdue University, West Lafayette, Indiana 47907, USA}
 
\author{F. Robicheaux}%
 \email{robichf@purdue.edu}
 \affiliation{Department of Physics and Astronomy, Purdue University, West Lafayette, Indiana 47907, USA}
 \affiliation{Purdue Quantum Science and Engineering Institute, Purdue University, West Lafayette, Indiana 47907, USA}

\date{\today}

\begin{abstract}
In recent years, chiral quantum optics has emerged as an active research area due to the promising applications in quantum information processing as well as nonlinear optics. In this work, we present results on the properties of the incoherent component of the transmitted light after transmons chirally coupled to a waveguide. The well-known resonance fluorescence of one chiral transmon resembles nonlinear quantum optics as it can convert the coherent light, with photons having spatially extensive coherence and Poisson distribution, into a field with spatially localized coherence and bunching photon statistics. However, another chiral transmon can undo the effect of the first transmon regardless of the Rabi frequency under the idealization involved in this work. A full wavefunction calculation shows that, in the weak driving limit, this incoherent light mainly comes from two-photon processes - one chiral transmon converts two independent photons into a momentum-correlated photon pair. In addition to the driving through the waveguide, we study the role of a local drive with independently tunable amplitude and phase that can address each transmon individually. The interplay between the waveguide and local driving modulates the contribution of the incoherent transmission and hence enables the engineering of the quantum statistics of the transmitted field, with tunable $g^{(2)}(0)$ spanning anti-bunched, coherent, and strongly-bunched regimes.
\end{abstract}

\maketitle


\section{\label{sec:introduction}Introduction}



Chiral quantum optics~\cite{Lodahl2017,PRXQuantum.6.020101}, where a chain of quantum emitters are coupled to a one-dimensional (1D) photonic continuum such that the emitter-photon interaction depends on the direction of photon propagation, paves the way to a new light-matter interface. Recent years have witnessed various techniques to implement chiral quantum systems. One common method is based on spin-momentum locking, where the transverse photon spin is opposite for opposite propagation directions~\cite{Lodahl2017,PRXQuantum.6.020101,BLIOKH20151,Bliokh2015,Aiello2015,Sinev2017,Francisco2013,Spitzer2018}. As a result, emitters with polarization-selective transitions undergo different scattering processes when coupled to light propagating in opposite directions. Another approach exploits the quantum nonlinearity of two-level systems. Here, the constructive or destructive interference between two pathways populating the quasi-dark state depends on the direction of the incoming light, resulting in a non-reciprocal scattering process~\cite{PhysRevLett.113.243601,PhysRevLett.121.123601}. Besides, two quantum emitters symmetrically coupled to a one-dimensional waveguide can interfere with each other such that their collective light-matter interaction depends on the propagation direction of the light~\cite{PhysRevA.102.053720,Guimond2020,Kannan2023,Redchenko2023}. Recently, chirality is also implemented in giant atoms, where a quantum emitter can be parametrically coupled to a one-dimensional waveguide at more than one point. The interplay between the phase difference of the couplings and the propagation phase of the photon gives rise to constructive/destructive interference between the photon emitted at different coupling points~\cite{PhysRevX.13.021039}. Experimentally, the chiral interaction has been implemented on platforms like neutral atoms~\cite{PhysRevLett.110.213604,Itay2014,Mitsch2014}, quantum dots~\cite{PhysRevLett.110.037402,Luxmoore2013,Sollner2015,Coles2016}, nanoparticles~\cite{Neugebauer2014,Jan2014}, and plasmonic surfaces~\cite{PhysRevLett.108.213907,Jiao2013,Francisco2013,OConnor2014}.

In recent years, artificial atoms realized by superconducting qubits emerge as a promising platform for quantum information processing~\cite{Roth2023,PhysRevA.69.062320,PhysRevA.75.032329,GU20171,Krantz2019}. Compared to other platforms, superconducting qubits exhibit very high coupling efficiency to a one-dimensional waveguide that reaches $99.9\%$~\cite{RevModPhys.95.015002,Mirhosseini2019}, allowing for efficient interaction in the quantum regime~\cite{Wallraff2004}. Besides, the confinement of the field to one dimension enables photons to propagate with negligible loss, making long-range interaction accessible~\cite{Arjan2013}. In addition, the possibility of individual control of each qubit offers another degree of control of the system~\cite{Ma2019,Botao2026}. Recent advances have demonstrated chiral couplings between superconducting qubits and one-dimensional waveguides ~\cite{PhysRevA.102.053720,Guimond2020,Kannan2023,Redchenko2023,PhysRevX.13.021039,PhysRevLett.121.123601,Almanakly2025}.

The chirality enables promising applications in quantum information processing like deterministic quantum state transfer~\cite{PhysRevLett.78.3221,Kurpiers2018,Northup2014,Axline2018,PhysRevLett.120.200501,PhysRevApplied.12.044067,PhysRevLett.125.260502,PhysRevLett.118.133601,PhysRevX.7.011035,PhysRevResearch.2.013369} or directional excitation transport~\cite{Yu_2026,PhysRevA.103.063711,Jen_2019}. Besides, the chiral interaction facilitates unique many-body dynamics in driven-dissipative systems such that even for this open system, the non-equilibrium steady state is a pure dark state~\cite{Stannigel_2012,PhysRevLett.113.237203,PhysRevA.91.042116}. 

Despite extensive efforts to create entanglement among quantum emitters, the effect of chirality on the light is less explored. In ~\cite{PhysRevA.107.013717}, a chain of chiral emitters under coherent drive are reported to generate a displaced mixture of Fock states, with nonclassicality characterized by the negativity in the Wigner function. Also, Ref.~\cite{PhysRevX.10.031011} reports the generation of photon bound states after propagating through chiral emitters. In Ref.~\cite{Prasad2020}, a chain of atoms chirally coupled to a nanofibre are employed to generate correlated photon states via resonance fluorescence - despite the weak atom-nanofibre coupling and photon loss, the cooperative enhancement of nonlinear interaction allows the observation of strong photon correlation (bunching or anti-bunching) depending on the atom number.

In this work, we consider transmons, a kind of superconducting qubits with less sensitivity to charge noise, chirally coupled to a one-dimensional waveguide. By sending classical/quantum light into the waveguide, we investigate the state of the light after passing through the transmon chain. Throughout this work, the transmitted field is divided into coherent and incoherent components - the coherent component can interfere with the incident field while the incoherent component cannot. We find that a single chiral transmon can convert coherent light, with photons having long-range coherence and Poisson distribution, into an incoherent field with spatially localized coherence and bunching photon statistics. This behavior is reminiscent of nonlinear quantum optics mediated by one chiral transmon. Remarkably, another chiral transmon can transform the incoherent light back to a coherent field regardless of the power of the incident light. For weak light, the full quantum calculation identifies a two-photon process as the main origin of this incoherent light, where a single chiral transmon generates a photon pair with opposite momentum shift out of two independent photons. The wavefunction of this photon pair is orthogonal to the incoming wavepacket and another chiral transmon suppresses its population, which accounts for the loss/revival of coherence after one/two chiral transmon(s). Finally, we introduce a local driving with individually tunable amplitude and phase on each transmon in addition to the driving through the waveguide. The interplay between these two kinds of driving can alter the contribution of the incoherent transmission after one chiral transmon and hence generates photon states with tunable statistics.

This work is organized as follows: Sec.~\ref{sec:sys_setup} gives an overview of the system setup, Sec.~\ref{sec:loss_and_revive_of_coherence} describes the equations of motion for semi-classical calculations and corresponding results. Sec.~\ref{sec:two_photon_process} covers the dynamics and results of the full-quantum calculations. Sec.~\ref{sec:engineered_driving} explores the interplay between individual local driving and the waveguide driving. The summary and outlook are given in Sec.~\ref{sec:summary}.

\section{System setup\label{sec:sys_setup}}

The system under investigation consists of a chain of $N$ transmons chirally coupled to a one-dimensional waveguide, as is shown in Fig.~\ref{fig:semi-classical}. The chiral coupling means the transmons only couple to the right-propagating photon. In this work, the transmons are considered as two-level systems, with the ground state denoted by $\ket{0}$ and the excited state represented by $\ket{1}$. The related Pauli spin operators are:

\begin{equation}
    \hat{\sigma}_n^+ \!=\! \ket{1}_n\!\bra{0}_n\!, \  \hat{\sigma}_n^- \!=\! \ket{0}_n\!\bra{1}_n\!,\  \hat{\sigma}_n^z \!=\! \ket{1}_n\!\bra{1}_n \!-\! \ket{0}_n\!\bra{0}_n,
\end{equation}
where $n$ is the index of the transmon. The simulation with the photon part described by a classical continuous wave is called ``semi-classical'' simulation, as described in Sec.~\ref{sec:loss_and_revive_of_coherence}. In contrast, if the calculation involves the wavefunction representation of the photon part, we call it a ``full quantum'' calculation, which will be elaborated in Sec.~\ref{sec:two_photon_process}. Section~\ref{sec:engineered_driving} explores the interplay between local and waveguide driving. Throughout this work, we assume $\hbar = 1$ and all the transmons are perfectly coupled to the waveguide, corresponding to a waveguide coupling efficiency of unity. 
Unless otherwise stated, the parameters used in this work are: $\Gamma = 2\pi \times 10^6 s^{-1}$, $\omega_a = 2 \pi \times 5 \times 10^9 s^{-1}$, $v_g = 2 \times 10^8 m/s$, $k_0 = \omega_a / v_g = 2 \pi \times 25 \ m^{-1}$, and the range of momentum $\abs{k - k_0} \le 15.7 \Gamma/v_g$.

\section{Semi-classical simulations: Loss and Revival of Coherence \label{sec:loss_and_revive_of_coherence}}

This section describes the dynamics of the system under coherent-state driving at a fixed momentum $k$ and reviews some results from resonance fluorescence. Some results in this section have been discussed in several works, for example~\cite{PhysRevX.13.021039,RevModPhys.95.015002}.
\begin{figure}[!htbp]
  \centering
  \includegraphics[width = 0.45\textwidth]{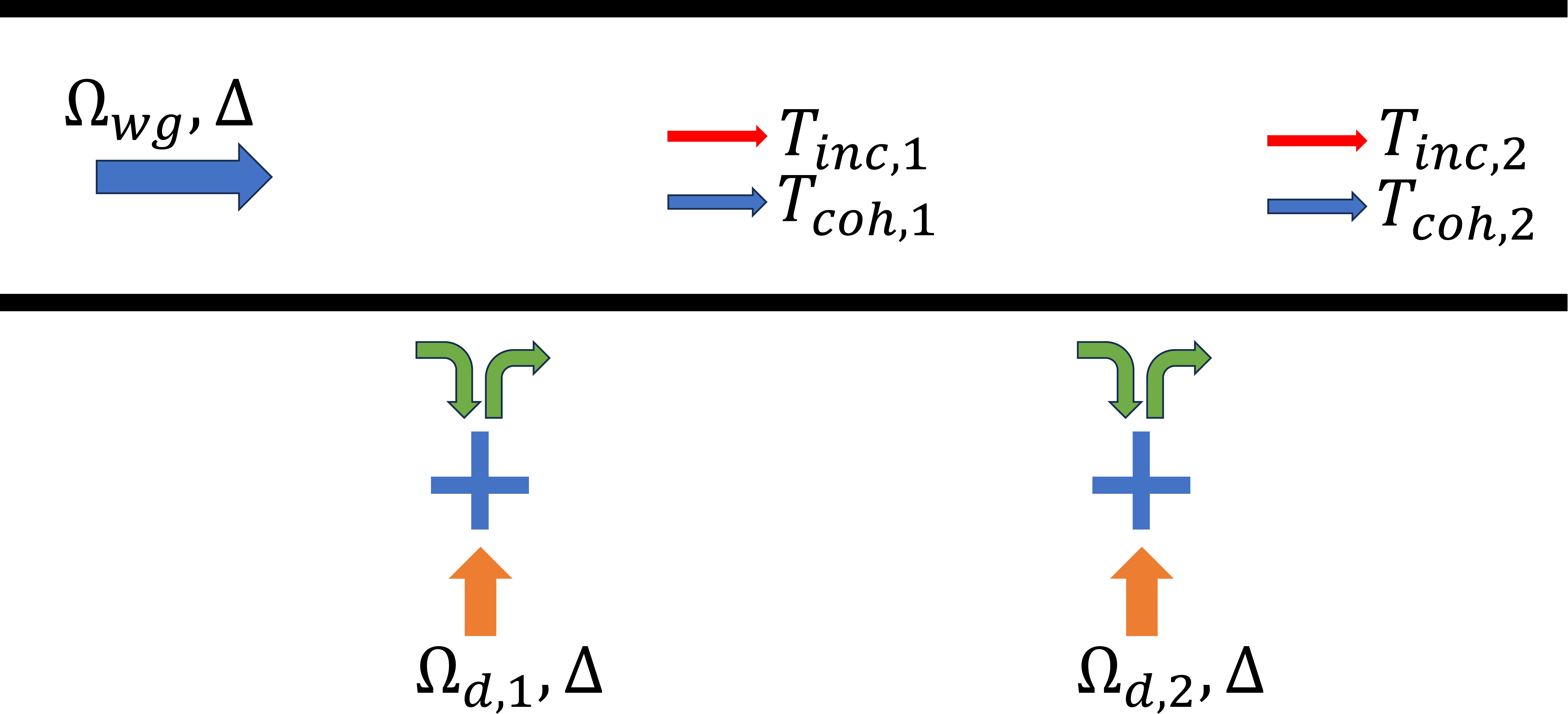}
  \caption{Schematics of the system: Two transmons chirally coupled to a 1D waveguide is driven by two continuous-wave fields, $\Omega_{wg}$ through the waveguide and $\Omega_d$ by the side port, with the same detuning. }
  \label{fig:semi-classical}
\end{figure}

The schematics of the system are shown in Fig.~\ref{fig:semi-classical}. In this case, the photon part is considered as a coherent state and, by the Born-Markov approximation, can be traced out, leaving the reduced density operator of the transmons, $\hat{\rho}$, and the dynamics dominated by the master equation~\cite{PhysRevA.95.033818,PhysRevResearch.2.043213,PhysRevResearch.3.033233}:

\begin{equation}
    \frac{d}{dt} \hat{\rho} = -i[H, \hat{\rho}] + \mathcal{L}[\hat{\rho}].
\end{equation}
The Hamiltonian consists of two parts:
\begin{equation}
    H = H_l + H_{wg},
\end{equation}
where $H_l$ corresponds to the coherent driving using the rotating wave approximation:
\begin{equation}
     H_l = \sum_n \left (\frac{\Omega_n}{2} \hat{\sigma}_n^+ + \frac{\Omega_n^*}{2} \hat{\sigma}_n^- - \frac{\Delta_n}{2} \hat{\sigma}_n^z \right ),
    \label{eq:H_l}
\end{equation}
where $\Omega_n$ is the Rabi frequency of transmon $n$ and $\Delta_n$ is the corresponding detuning:
\begin{equation}
    \Delta_n = \omega_l - \omega_{a, n},
\end{equation}
where $\omega_l$ is the frequency of the incoming light and $\omega_{a, n}$ is the transition frequency of transmon $n$. As is shown in Fig.~\ref{fig:semi-classical}, the overall coherent drive consists of two components with the same detuning:
\begin{equation}
    \Omega_n = \Omega_{wg, n} + \Omega_{d, n},
    \label{eq:overall_rabi_freq}
\end{equation}
where $\Omega_{wg, n}$ is the Rabi frequency from the waveguide driving, with a phase determined by the position of the transmon, $x_n$, and the wave number of the incident mode, $k$:
\begin{equation}
    \Omega_{wg, n} = \abs{\Omega_{wg, n}} e^{ikx_n},
\end{equation}
and a local drive from the side port, with independently tunable phase and amplitude:
\begin{equation}
    \Omega_{d, n} = \abs{\Omega_{d, n}} e^{i\phi_{n, d}}.
\end{equation}
The chiral exchange interaction between transmons via the waveguide is described by $H_{wg}$:
\begin{equation}
    H_{wg} = -i\frac{\Gamma}{2}\sum_{n > m} (e^{ik\abs{x_n - x_m}} \hat{\sigma}_n^+ \hat{\sigma}_m^- - H.c.).
    \label{eq:H_wg}
\end{equation}
The dissipative process is described by the Lindblad superoperator:
\begin{equation}
    \mathcal{L}[\hat{\rho}] = \sum_{m, n}\frac{\Gamma_{mn}}{2}(2\hat{\sigma}_m^- \hat{\rho} \hat{\sigma}_n^+ - \hat{\sigma}_n^+ \hat{\sigma}_m^- \hat{\rho} - \hat{\rho} \hat{\sigma}_n^+ \hat{\sigma}_m^-),
    \label{eq:L}
\end{equation}
where $\Gamma_{mn} = \Gamma e^{-ik(x_m - x_n)}$.

We start all the transmons in the ground state and introduce a constant coherent drive through the waveguide so that the system reaches a steady state, $\hat{\rho}(\infty)$. Out of all the transmitted light, one question is how much can interfere with the incoming light. This quantity, referred as coherent transmission, can be evaluated by the input-output theory of waveguide QED~\cite{PhysRevA.88.043806}:

\begin{equation}
    \hat{\alpha} = \mathbbm{1} - i \frac{2\Gamma}{\Omega_{wg}} \sum_n e^{-ikx_n} \hat{\sigma}^-_n,
    \label{eq:alpha}
\end{equation}
where $\hat{\alpha}$ is the output field operator whose expectation value is the transmission amplitude, $t_{coh}$. The identity operator corresponds to the coherent incoming light and the second term refers to the contribution from the transmon emission. In this section, we consider the case with only waveguide driving, where the overall Rabi frequency $\Omega = \Omega_{wg}$. The coherent transmission probability, $T_{coh}$, is defined as its amplitude square:

\begin{equation}
    t_{coh} = \langle \hat{\alpha} \rangle, \quad T_{coh} = \abs{\langle \hat{\alpha} \rangle}^2.
    \label{eq:Tcoh}
\end{equation}
The incoherent transmission, as the complementary component of the transmitted light, can be obtained by subtracting the coherent transmission from the total transmission:

\begin{equation}
    T_{inc} = T_{tot} - T_{coh},
    \label{eq:Tinc}
\end{equation}
where the total transmission probability:

\begin{equation}
    T_{tot} = \langle \hat{\alpha}^\dagger \hat{\alpha} \rangle
    \label{eq:Ttot}
\end{equation}
is guaranteed to be $1$ because we only introduce the coherent drive through the waveguide in this section. 

\begin{figure}[!htbp]
  \centering
  \includegraphics[width = 0.5\textwidth]{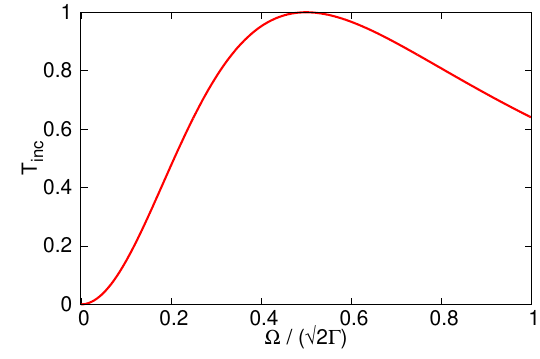}
  \caption{On-resonance incoherent transmission probability after one chiral transmon as a function of normalized Rabi frequency, corresponding to the analytical expression from Eq.~\eqref{eq:Tinc_Omgswp}.}
  \label{fig:Tinc_Omgswp}
\end{figure}

For the interest of this work, Fig.~\ref{fig:Tinc_Omgswp} reviews some results of the resonance fluorescence of chiral quantum emitters~\cite{PhysRevX.13.021039}, demonstrating the analytical expression of on-resonance $T_{inc}$ in terms of the normalized Rabi frequency $\Omega/\Gamma$, which reads:

\begin{equation}
    T_{inc} = 1 -  \abs{1 - i \frac{2\Gamma}{\Omega} \langle \hat{\sigma}^-_1 \rangle } ^ 2 = \frac{\frac{\abs{\Omega}^2}{2} \Gamma ^ 2}{(\Delta^2  + \frac{\abs{\Omega}^2}{2} + \frac{\Gamma^2 }{4}) ^ 2},
    \label{eq:Tinc_Omgswp}
\end{equation}
where
\begin{equation}
    \langle \hat{\sigma}^-_1 \rangle = \frac{\frac{\Omega}{2} (- \Delta - i\frac{\Gamma}{2})}{\Delta^2  + \frac{\abs{\Omega}^2}{2} + \frac{\Gamma^2 }{4}}.
\end{equation}

Note that $T_{inc}$ grows as $\Omega$ increases initially but drops when passing $\Omega = 0.5\sqrt{2}\Gamma$, at which $T_{inc} = 1$. This means all the transmitted field cannot interfere with the incoming light and this phenomenon is considered as "loss of coherence" in this work. At the special point ($\Omega = 0.5\sqrt{2} \Gamma$), one can check that the incoming laser (corresponding to the $1$ in the bracket of Eq.~\eqref{eq:Tinc_Omgswp}) exactly cancels the contribution from the transmon emission (term $i {2\Gamma}\langle \hat{\sigma}^-_1 \rangle/{\Omega} $), leading to $t_{coh} = 0$ and hence the loss of coherence.

\begin{figure}[!htbp]
  \centering
  \includegraphics[width = 0.5\textwidth]{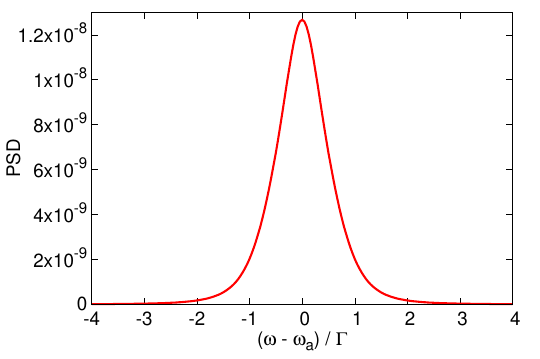}
  \caption{Power spectrum density of the incoherent component as a function of the normalized detuning at $\Omega = 0.5\sqrt{2}\Gamma$, where the coherence of transmitted light is completely lost.}
  \label{fig:psd_wg}
\end{figure}
\begin{figure}[!htbp]
  \centering
  \includegraphics[width = 0.5\textwidth]{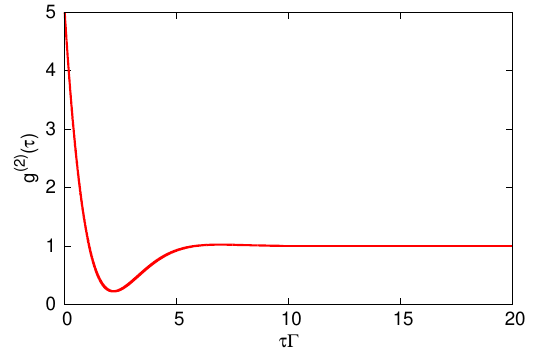}
  \caption{Second-order correlation function of the incoherent component for $\Omega = 0.5\sqrt{2}\Gamma$, where $T_{coh} = 0$. }
  \label{fig:g2_wg}
\end{figure}

The properties of this incoherent light are characterized by the power spectrum density and photon statistics (See Appendix~\ref{app:g1_g2}). 
As a benchmark of the resonance fluorescence of one chiral emitter\cite{PhysRevX.13.021039}, Fig.~\ref{fig:psd_wg} illustrates the power spectrum density of the incoherent component of transmission at $\Omega = 0.5\sqrt{2} \Gamma$, while Fig.~\ref{fig:g2_wg} demonstrates the corresponding $g^{(2)}$ function. The broadening in Fig.~\ref{fig:psd_wg} indicates the spatially localized coherence while $g^{(2)}(0) = 5 > 1$ in Fig.~\ref{fig:g2_wg} suggests the bunching behavior. 
This implies that the chiral transmon can convert coherent light, with photons having long-range coherence and Poisson distribution, into light with spatially localized coherence and bunching photon statistics, which is reminiscent of the nonlinear optics mediated by the chiral transmon.

\begin{figure}[!htbp]
  \centering
  \includegraphics[width = 0.5\textwidth]{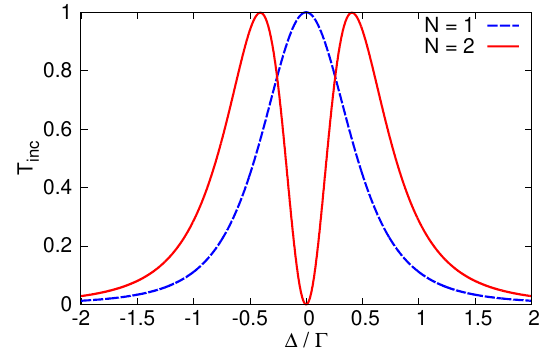}
  \caption{Incoherent transmission after one (blue dashed line) and two (red solid line) chiral transmons as a function of normalized detuning at $\Omega = 0.5 \sqrt{2} \Gamma$. For zero detuning, there is complete loss of coherence after one transmon and complete restoration of coherence after two transmons.}
  \label{fig:Tinc_Nbit2}
\end{figure}

Another motivation for investigating this incoherent component stems from the phenomenon "revival of coherence." As shown in Fig.~\ref{fig:Tinc_Nbit2}, the on-resonance $T_{inc} = 1$ after one chiral transmon but $T_{inc} = 0$ after two transmons. This means the second chiral transmon can undo the effect of the first chiral transmon (broadening of power spectrum and bunching photon statistics) and convert the incoherent light back to coherent light. One can show that the two transmons form a pure dark steady state, commonly referred to as a ``dimer''\cite{Stannigel_2012,PhysRevLett.113.237203,PhysRevA.91.042116}. In this state, they are effectively decoupled from the waveguide field, leading to coherent transmission. This also rules out a completely random phase of the incoherent transmitted light after a single chiral transmon as the origin of the loss of coherence. To further understand which process transforms coherent light into a field with spatially localized coherence and bunching photon statistics, and to uncover the reason for the lack of interference, full information about the light field is necessary. This motivates the full-quantum simulations that explicitly include the photon part in the wavefunction.

\section{Full-quantum calculations: Two-photon Processes\label{sec:two_photon_process}}

\begin{figure}[!htbp]
  \centering
  \includegraphics[width = 0.45\textwidth]{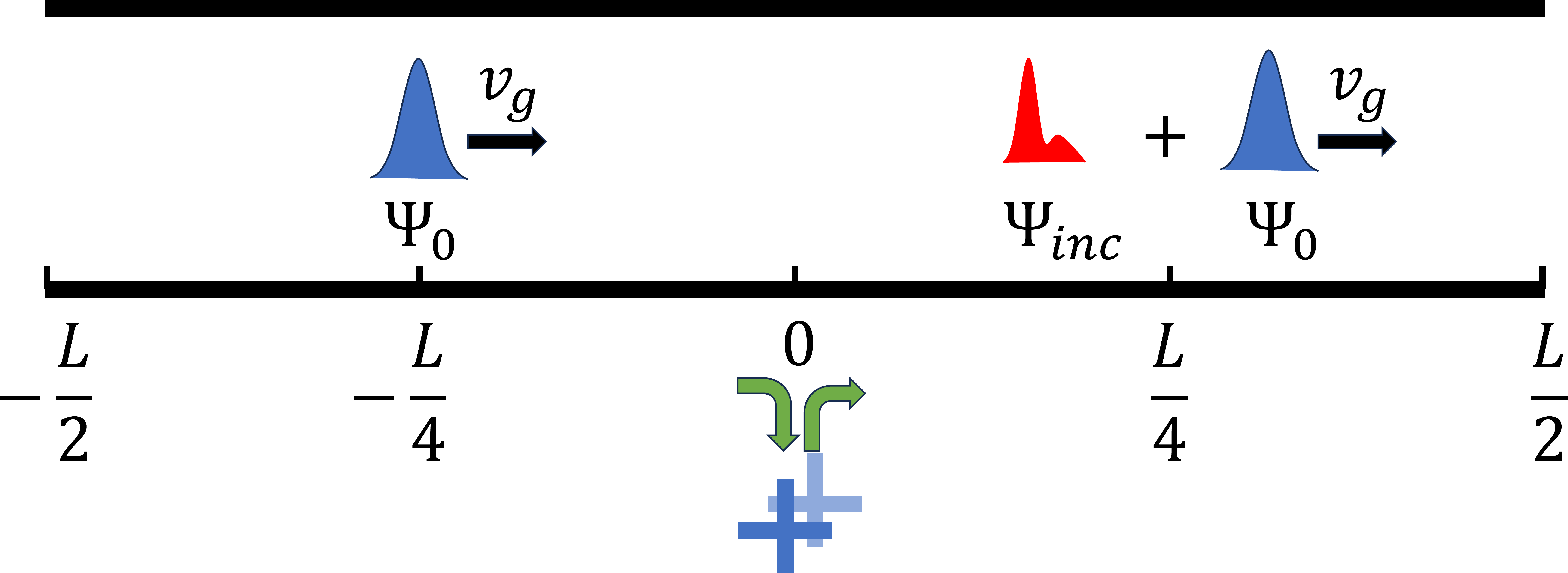}
  \caption{Schematics of the full-quantum calculation: One (two) transmon(s) chirally coupled to a 1D waveguide, with a two-photon wavepacket $\Psi_0$ centered at $x = -L/4$ propagating to the right with group velocity $v_g$. At final time, the wavefunction is centered at $x = L/4$ and is divided into two components $\Psi_0$ and $\Psi_{inc}$.}
  \label{fig:full-quantum}
\end{figure}

This section investigates the system dynamics using a full-quantum approach and benchmarks the results against semi-classical calculations in the weak-driving regime. As is shown in Fig.~\ref{fig:full-quantum}, we consider a one-dimensional dispersionless waveguide with length $L$ and a right-propagating two-photon wavepacket with group velocity $v_g$. As described below, the wave packet does not reach the end of the waveguide.
Under periodic boundary condition, the resolution in the momentum space is $2\pi/L$ and the momentum $k$ takes the values:
\begin{equation}
    k_m =  m \frac{2\pi}{L}, \quad m = ..., -2, -1, 0, 1, 2, ...
\end{equation}
To study the interaction process in the quantum regime, we represent the state of the system by a wavefunction in the two-excitation subspace taking both the transmons and the photons into account.
\begin{equation}
\begin{split}
    \ket{\Psi(t)} &= [\sum_{n > n^\prime} \psi_2(n, n^\prime, t)\hat{\sigma}_n^+ \hat{\sigma}_{n^\prime}^+ + \sum_{n, m} \psi_1(n, k_m, t) \hat{a}_m^\dagger \hat{\sigma}_n^+ \\
    &+\!\!\! \sum_{m \ge m^\prime \ge 0}\Psi(k_m, k_{m^\prime}, t) / \sqrt{1 + \delta_{m, m^\prime}} \hat{a}_m^\dagger \hat{a}_{m^\prime}^\dagger] \ket{0, vac},
\end{split}
\label{eq:ket_Psi}
\end{equation}
where $\ket{0, vac}$ is the ground state for both transmons and the photon part. $\psi_2(n, n^\prime, t)$ represents the amplitude for transmons $n$ and $n^\prime$ being simultaneously excited. $\psi_1(n, k_m, t)$ describes a state with one excited transmon $n$ and one photon of momentum $k_m$, whereas $\Psi(k_m, k_{m^\prime}, t)$ represents a two-photon state with momenta $k_m$ and $k_{m^\prime}$.
The dynamics of the system arises from the Hamiltonian~\cite{PhysRevA.88.043806,PhysRevLett.95.213001,PhysRevLett.98.153003,PhysRevA.82.063816}: 
\begin{equation}
    H = H_A + H_F + H_{AF},
\end{equation}
where $H_A$ is the energy of the transmon part:
\begin{equation}
    H_A = \sum_n \omega_a \hat{\sigma}_n^+\hat{\sigma}_n^-,
\end{equation}
while $H_F$ is the energy of the photon part:
\begin{equation}
    H_F = \sum_m \omega_m \hat{a}_m^\dagger \hat{a}_m.
\end{equation}
where $\hat{a}_m^\dagger$ is the creation operator of a photon with momentum $k_m$ and energy $\omega_m$, which satisfies the canonical commutation relation:
\begin{equation}
    [\hat{a}_m, \hat{a}_{m^\prime}^\dagger] = \delta_{m, m^\prime}.
\end{equation}
After the rotating wave approximation, the coupling term reads:
\begin{equation}
    H_{AF} = \sum_n \sum_{m > 0} (g_n e^{-i k_m x_n} \hat{a}_m^\dagger \hat{\sigma}_n^- + h.c.),
\end{equation}
where $g_n$ is the coupling at transmon $n$:
\begin{equation}
    g_n = \sqrt{\Gamma_n\frac{v_g}{L}},
\end{equation}
where $\Gamma_n$ is the decay rate of individual transmon $n$. The chirality manifests as the summation over the positive $k$. This full-quantum calculation inherently accounts for the non-Markovian effects. To mimic the case of semi-classical simulation with Markov approximation, we assume that all the transmons are the same and are located at nearly the same position $x_{n+1} - x_n \ll \lambda$ so that the coupling is not dependent on $k$, that is, $g_n e^{-i k_m x_n} = g = \sqrt{\Gamma\frac{v_g}{L}}$.

\begin{figure}[!htbp]
  \centering
  \includegraphics[width = 0.5\textwidth]{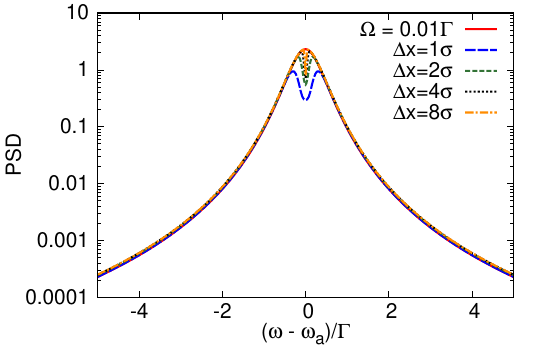}
  \caption{Comparison between the power spectrum densities from full quantum calculation and semi-classical simulations.
  For semi-classical simulations, we use the incoming light with $\Omega = 0.01\Gamma$ and $\Delta = 0$. The full quantum calculations involve a two-photon Gaussian wavepacket, and the four curves correspond to different wavepacket widths $\Delta x$.}
  \label{fig:psd}
\end{figure}

In this notation, the equations of motion are described by the Schrodinger's equation:
\begin{equation}
    i\frac{d}{dt}\ket{\Psi} = H\ket{\Psi}.
    \label{eq:dPsi_dt}
\end{equation}
We initialize the wavefunction to be fully in the two photon subspace with $\Psi (x_1,x_2,0)$ equivalent to a product of two Gaussian wavepackets with width $\Delta x = L/24$:
\begin{equation}
    \Psi_0(x_1, x_2, 0) = \frac{1}{\sqrt{2\pi} \Delta x} e^{-\frac{(x_1 - x_c)^2 + (x_2 - x_c)^2}{4\Delta x^2}+ ik_0 (x_1 + x_2)},
    \label{eq:psi_0_x1_x2}
\end{equation}
where $k_0$ is the momentum on resonance:
\begin{equation}
    k_0 = \omega_a / v_g,
\end{equation}
and $x_c(t)$ is the center of the wavepacket, which is initialized at $x_c(0) = -L/4$ and moves at group velocity $v_g$. The transmons located at $x = 0$ are in the ground state initially. We time integrate Eq.~\ref{eq:dPsi_dt} until the wavepacket arrives at $x_c(t_f) = L/4$ at $t = t_f = (L/2)/v_g$. For the parameters in our calculations, the probability for a transmon to be excited is less than $10^{-16}$, rendering the final wavefunction $\Psi(x_1, x_2, t_f)$ in real space and $\Psi(k, k^\prime, t_f)$ in momentum space. This wavefunction has a finite component from the trivial displacement of the initial photon, denoted by $\Psi_0(k, k^\prime, t_f)$: 
\begin{equation}
    \Psi_0(k, k^\prime, t_f) = \frac{2\Delta x}{\sqrt{2\pi}}e^{-\Delta x^2 [(k - k_0)^2 + (k^\prime - k_0)^2] - i(k + k^\prime) x_c},
\end{equation}
By definition, this component can interfere with itself, i.e., the incoming photon after free propagation, and is considered as the coherent component. To find the incoherent component for comparison with the semi-classical results in the weak-driving regime, an orthogonalization process can be used to remove $\Psi_0$:
\begin{equation}
    \ket{\Psi_{inc}} = \ket{\Psi} - \ket{\Psi_0}\bra{\Psi_0}\ket{\Psi}.
    \label{eq:Psi_inc}
\end{equation}
Note that $\Psi(k_m, k_{m^\prime}, t)$ in Eq.~\eqref{eq:ket_Psi} is only defined to be nonzero when $m \ge m^\prime$. For the purpose of plotting, we symmetrize $\Psi_{inc}$ by:
\begin{equation}
    \Psi_{inc}^{sym}(k, k^\prime) = \frac{1}{\sqrt{2}} [\Psi_{inc}(k, k^\prime) + \Psi_{inc}(k^\prime, k)]/\sqrt{1 + \delta_{k, k^\prime}}.
    \label{eq:sym_wavefunction}
\end{equation}
Hereafter, all the wavefunction-related observables are calculated using $\Psi_{inc}^{sym}$ and the Lamb shift is ignored unless otherwise mentioned. The power spectrum density can be obtained by the Fourier transform of the auto-correlation function of $\Psi_{inc}^{sym}(x_1, x_2, t_f)$ and takes the form:

\begin{equation}
    \begin{split}
        S(k) &= N_S \int_0^{+\infty} \abs{\Psi_{inc}^{sym}(k, k^\prime, t_f)}^2 dk^\prime
    \end{split}
    \label{eq:psd_wavepacket}
\end{equation}
where $N_S$ is the normalization factor and $k$ can be transformed into $\omega$ by the dispersion relation of the waveguide $\omega = v_g k$. With a fixed photon number, increasing the spatial width of the wavepacket narrows its momentum distribution, causing it to approach the continuous-wave limit in the weak-driving regime. In Fig.~\ref{fig:psd}, we plot the power spectral densities for four $\Psi_0$ with $\Delta x$ in different multiples of $\sigma = 20v_g / (3\sqrt{2}\Gamma)$ and compare them with the semiclassical results at $\Omega = 0.01\Gamma$ to examine their convergence toward the semiclassical prediction in the weak-driving regime. The length of the waveguide $L$ takes the values $160 v_g / \Gamma, \ 320v_g/\Gamma, \ 640v_g/\Gamma, \ 1280 v_g/\Gamma$, corresponding to the curves with $\Delta x = 1\sigma, \ 2\sigma, \ 4\sigma, \ 8\sigma$.

Numerical results show that the contribution of $\Psi_{inc}^{sym}$ in $\Psi^{sym}$, evaluated by the ratio $\abs{\Psi_{inc}^{sym}}^2 / \abs{\Psi^{sym}}^2$, scales as $\sim 1/\Delta x$ after one chiral transmon, rendering the power spectrum of $\Psi_{inc}^{sym}$ after one chiral transmon scaling as $\sim 1/\Delta x$. To compensate for this effect, we introduce the normalization factor $N_S \propto \Delta x$ to the power spectrum after one chiral transmon so that one can focus on the comparison between the shape of the functions. One can check that by increasing $\Delta x$, the power spectrum density of the full quantum calculations is approaching the semi-classical results. The dip in the middle results from the finite sampling range as well as the varying profile of the Gaussian wavepacket and is shrinking with increasing $\Delta x$. Doubling the range of momentum does not significantly change the power spectrum density: The largest relative error between the two ranges appears at the end of $k$ range, where the values are smaller by $5$ orders of magnitude compared to the peak. Meanwhile, the relative error is about $1\%$ at the peak.

\begin{figure}[!htbp]
  \centering
  \includegraphics[width = 0.5\textwidth]{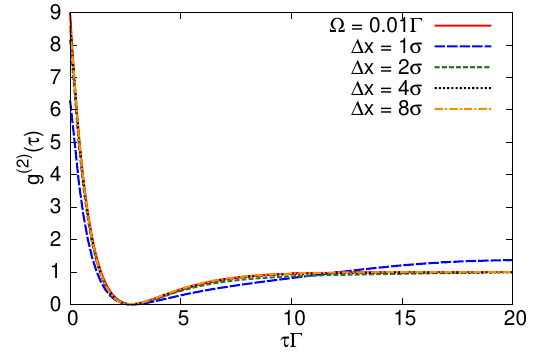}
  \caption{Comparison between the $g^{(2)}$ function from full quantum calculation and semi-classical simulations. The parameters are the same as Fig.~\ref{fig:psd}.}
  \label{fig:g2}
\end{figure}

As stated in Eq.~\eqref{eq:psd_wavepacket}, the power spectrum density is proportional to a marginal distribution of the wavefunction amplitude square and cannot uniquely determine the two-photon wavefunction. Therefore, having the same power spectral density does not necessarily imply that the incoherent field from the semi-classical result and the two-photon wavepacket are identical. For further verification, we compare the intensity correlation function of $\Psi^{sym}$ (the total transmitted wavefunction symmetrized similarly to Eq.~\eqref{eq:sym_wavefunction}) having different $\Delta x$ with the semi-classical results. The wavefunction $\Psi^{sym}(k, k^\prime, t_f)$ can be converted into real space by Fourier transform:

\begin{equation}
    \Psi^{sym}(x_1, x_2, t_f) = \mathcal{F}^{-1}[\Psi^{sym}(k, k^\prime, t_f)]
\end{equation}
and $x_1$, $x_2$ can be transformed into times $t_1$, $t_2$ by $x_i = v_g t_i$, leading to a two-time wavefunction $\Psi^{sym}(t_1, t_2, t_f)$ and the corresponding two-time probability $p_2(t_1, t_2, t_f) = \abs{\Psi^{sym}(t_1, t_2, t_f)}^2$. The one-time probability $p_1(t_i, t_f)$ can be obtained by $p_1(t_i, t_f) = \int p_2(t_i, t_j, t_f) dt_j$, where $i, j = 1, 2$ and $i \neq j$. Given a relative time $\tau = t_1 - t_2$, the $g^{(2)}$ function can be obtained by:

\begin{equation}
    g^{(2)}(\tau) = \frac{p_2(t_{max} + \tau/2, t_{max} - \tau/2, t_f)}{p_1(t_{max} + \tau/2, t_f)p_1(t_{max} - \tau/2, t_f)},
    \label{eq:g2}
\end{equation}
where $t_{max}$ is the time when the probability of $\Psi_0$, the final wavefunction after free propagation, reaches its peak.
The comparison of the $g^{(2)}$ function between the semi-classical results (see Appendix~\ref{app:g1_g2}) and the full quantum calculations is illustrated in Fig.~\ref{fig:g2}.

One can check that the $g^{(2)}$ functions become closer to the semi-classical results as the wavepackets have larger $\Delta x$. The largest error appears at $\tau = 0$ and does not converge to the semi-classical results for the range of momentum $\abs{k - k_0} \le 15.7 \Gamma/v_g$: $g^{(2)}(0) \approx 6.29, \ 8.16, \ 8.56, \ 8.55$ at $\Delta x = 1\sigma, \ 2\sigma, \ 4\sigma, \ 8\sigma$, respectively, leading a persistent relative error of $\sim5\%$. By doubling the range of $k$ from $15.7 \Gamma/v_g$ to $31.4 \Gamma/v_g$, the relative error further reduces to about $2\%$. This suggests that the error also results from the convergence with respect to the range of $k$. The convergence observed in Figs.~\ref{fig:g2} and~\ref{fig:psd} implies that the semiclassical and full quantum results exhibit asymptotically consistent properties of transmitted field as the input two-photon wavepacket approaches the weak coherent drive. Note that the incoherent transmission probability in Eq.~\eqref{eq:Tinc_Omgswp} $\propto \Omega^2$ in the weak-driving limit, and the power of the transmitted light is also $\propto \Omega^2$, leading to the power of the incoherent transmitted light $\propto \Omega^4$ - this provides evidence that the generation of the steady-state incoherent light from the weak coherent drive is mainly a two-photon process. Besides, this weak coherent drive ($\Omega = 0.01\Gamma$) exhibits photon flux on the order of $10^{-4}\Gamma$. So the $N_{ph}$-photon components with $N_{ph} > 2$ do not contribute substantially to the incident coherent field. Meanwhile, a one-photon process should preserve the power spectrum density due to energy conservation, rendering the Gaussian power spectrum of transmitted photons. However, the power spectral density in Fig.~\ref{fig:psd} is clearly non-Gaussian, ruling out the single-photon processes as the origin of the incoherent light we observe. So we can conclude that the incoherent component of the transmitted light after one chiral transmon mainly comes from a two-photon process for small $\Omega$. By Eq.~\eqref{eq:Psi_inc}, $\Psi_{inc}$ is orthogonal to the input wavepacket $\Psi_0$ and hence cannot interfere with it. This accounts for the emergence of incoherent transmission in the coherent driving case. Further, we observe the suppression of the incoherent component induced by another chiral transmon, with $\abs{\Psi_{inc}^{sym}}^2 / \abs{\Psi^{sym}}^2 \sim 1/\Delta x^3$. This means almost all of the incoherent component is transformed back to $\Psi_0$, which is consistent with the revival of coherence in the semi-classical results.

\begin{figure}[!htbp]
  \centering
  \includegraphics[width = 0.5\textwidth]{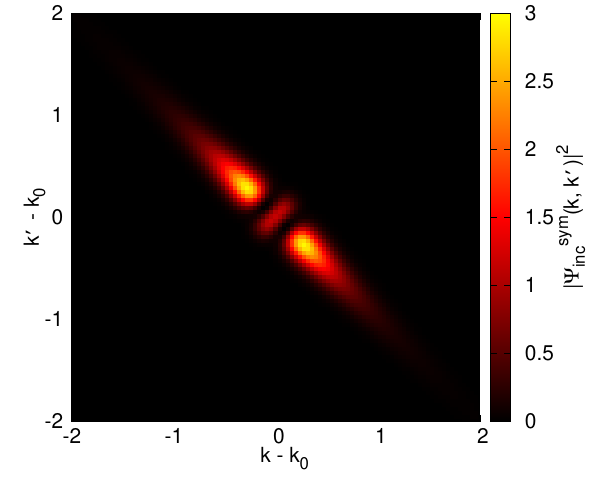}
  \caption{The two-photon density $\abs{\Psi_{inc}^{sym}(k, k^\prime, t_f)}^2$ as a function of $k - k_0$ and $k^\prime - k_0$ at $t_f$ for $\Delta x = 1\sigma$. Here the momenta $k - k_0$ and $k^\prime - k_0$ are in units of $\Gamma / v_g$.}
  \label{fig:two_photon_corr}
\end{figure}

For such localized and bunched two photons, a natural question is how these two photons are correlated. To illustrate this, we plot the two-photon probability density $\abs{\Psi_{inc}^{sym}(k, k^\prime, t_f)}^2$ as a function of $k - k_0$ and $k^\prime - k_0$ in Fig.~\ref{fig:two_photon_corr}.

Clearly, the probability density mainly locates along $k - k_0 = - (k^\prime - k_0)$. This implies the nonlinear quantum optics mediated by a single chiral transmon, which converts two independent photons into a photon pair with opposite momentum shift. By Eq.~\ref{eq:psd_wavepacket}, the power spectrum density results from the marginal distribution of $\abs{\Psi_{inc}^{sym}(k, k^\prime, t_f)}^2$. As a result, the finite spread of $\abs{\Psi_{inc}^{sym}(k, k^\prime, t_f)}^2$ on the line $k - k_0 = - (k^\prime - k_0)$ is projected to the broadening in the power spectrum. This is different from the broadening due to the uncertain emission time in spontaneous emission, which can happen in the one-photon interaction case. 

\begin{figure}[!htbp]
  \centering
  \includegraphics[width = 0.5\textwidth]{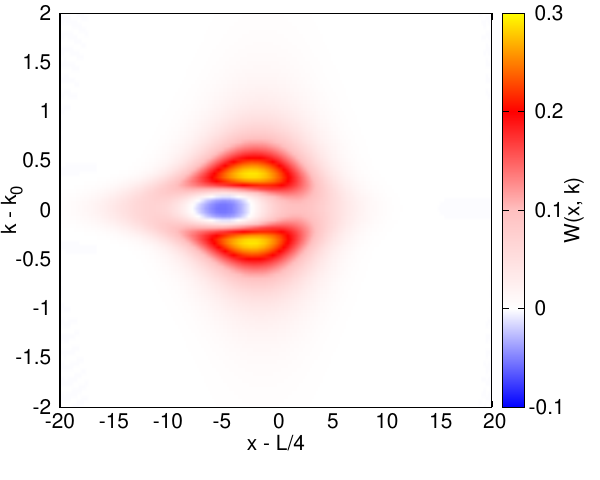}
  \caption{The one-photon Wigner function of the component $\Psi_{inc}^{sym}$, for the width of input pulse $\Delta x = 1\sigma$. The quadrature $x$ is in units of $v_g / \Gamma$ while $k$ is in units of $\Gamma / v_g$. }
  \label{fig:wigner}
\end{figure}

To check the nonclassicality of this anti-correlated photon pair, we trace out one photon and use the reduced density operator to calculate the one-photon Wigner function:
\begin{equation}
    W(x, k) = \frac{1}{2\pi}\int dk^\prime e^{-ik^\prime x} \rho(k - \frac{k^\prime}{2}, k + \frac{k^\prime}{2}),
\end{equation}
as is shown in Fig.~\ref{fig:wigner}.
The negativity in the middle suggests that the light is non-classical. Compared to the results in~\cite{PhysRevA.107.013717}, where the negativity of the Wigner function originates from the non-classical one-photon and two-photon contributions for driving strength comparable to $\Gamma$. In this work, we show that this negativity mainly comes from the two-photon process in the weak driving limit.

\section{Engineered driving: Generation of anti-bunching, coherent, bunching, and super-bunching light\label{sec:engineered_driving}}

\begin{figure}[!htbp]
  \centering
  \includegraphics[width = 0.5\textwidth]{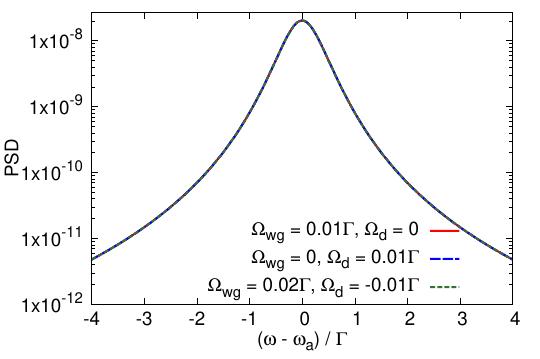}
  \caption{The power spectrum density of the incoherent transmission component for three combinations of $\Omega_{wg}$ and $\Omega_d$ while fixing $\Omega = 0.01\Gamma$. The curve for case (1) is normalized by multiplying the factor $f = 1250.27774349$ and the curve for case (2) is multiplied by $4f$.}
  \label{fig:psd_wg_sidedrv}
\end{figure}

In addition to the driving through the waveguide ($\Omega_{wg}$), transmons feature the flexibility of a local drive ($\Omega_d$) from a side port, with individually controllable amplitudes and phases on each transmon. This facilitates another degree of freedom so that one can manipulate $\Omega_{wg}$ and $\Omega_d$ to alter the coherent transmission amplitude in Eq.~\eqref{eq:alpha} while keeping the overall Rabi frequency $\Omega$ in Eq.~\eqref{eq:overall_rabi_freq} unchanged. 
This section considers only one transmon and exemplifies the interplay between $\Omega_{wg}$ and $\Omega_d$ with the same detuning $\Delta = 0$ while fixing the overall Rabi frequency $\Omega = 0.01\Gamma$. Especially, when the waveguide driving is absent, we remove the identity operator in Eq.~\eqref{eq:alpha} and scale the second term so that the field operator takes the form:
\begin{equation}
    \hat{\alpha} = -i \sum_n e^{-ikz_n} \hat{\sigma}^-_n.
\end{equation}
By definition, the $g^{(1)}(\tau)$ and $g^{(2)}(\tau)$ functions will not be affected by this scaling because the factors will cancel out in the evaluation, as elaborated in Appendix~\ref{app:g1_g2}. In this case, the equations of motion are exactly the same, except that the on-resonance coherent transmission amplitude after one chiral transmon should be generalized to:

\begin{equation}
    t_{coh} = 1 - \frac{\frac{\Gamma^2}{2} \frac{\Omega}{\Omega_{wg}}}{\frac{\abs{\Omega}^2}{2} + \frac{\Gamma^2}{4}}.
    \label{eq:Tcoh_wg_d}
\end{equation}
This coherent transmitted light has $g^{(2)}(0) = 1$ and cannot contribute to the bunching behavior. To obtain the bunched component, we need $t_{coh} = 0$, yielding the relation between the overall Rabi frequency and the waveguide Rabi frequency:

\begin{equation}
    \Omega_{wg} = \frac{\frac{\Gamma^2}{2}\Omega}{\frac{\abs{\Omega}^2}{2} + \frac{\Gamma^2}{4}}.
\end{equation}
In the weak driving limit, $\Omega \ll \Gamma$, one can check that $\Omega_{wg} \approx 2 \Omega$, implying that the local driving should have $\pi$ phase shift to the waveguide driving so that $\Omega_d = -\Omega = -0.5\Omega_{wg}$. To characterize this incoherent light, we plot the power spectrum density for three cases: (1) $\Omega_{wg} = \Omega = 0.01\Gamma$, (2) $\Omega_d = \Omega = 0.01\Gamma$, (3) $\Omega_d = -0.5 \Omega_{wg} = -\Omega = -0.01\Gamma$ in Fig.~\ref{fig:psd_wg_sidedrv}.

\begin{figure}[!htbp]
  \centering
  \includegraphics[width = 0.5\textwidth]{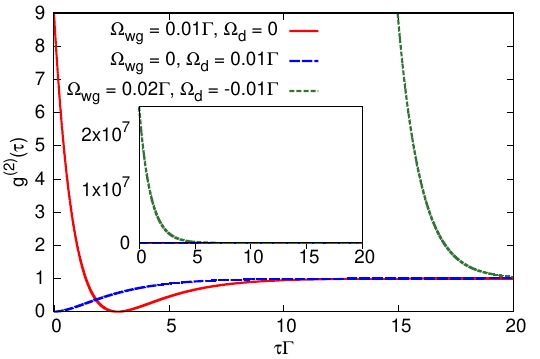}
  \caption{The $g^{(2)}$ function of the total transmitted field for three combinations of $\Omega_{wg}$ and $\Omega_d$. The inset is the zoom-out of the graph.}
  \label{fig:g2_wg_sidedrv}
\end{figure}

Note that the power spectrum densities are normalized for the same reason as Fig.~\ref{fig:psd}. The normalization factor for case (1) is $f = 1250.27774349$ while case (2) is multiplied by $4f$. This suggests that case (3), with only the incoherent transmission component, has the power much lower than the other two cases. In addition, case (1) and (2) also have the coherent component in general, leading to a much lower photon pair flux of case (3) compared to case (1) and (2). The agreement between the curves suggests that the power spectrum densities for these three cases are the same in terms of the shape. However, they do not necessarily correspond to the same light field: At least in case (2), the transmitted light should be anti-bunching because it only comes from the emission of the transmon, which cannot emit two photons simultaneously. To explicitly show this, we plot the $g^{(2)}$ function for the three cases in Fig.~\ref{fig:g2_wg_sidedrv}.

One can check that the three cases show very different photon statistics: (1) The case with only waveguide driving (red line) is highly bunching with $g^{(2)}(0) = 9$, (2) while the case with only local driving (blue dashed line) is anti-bunching with $g^{(2)}(0) = 0$, (3) and the case with both drivings so that the coherent transmission is $\sim 0$ gives a super-bunching light~\cite{Bhatti2015,Zeidan2026}, with $g^{(2)}(0) \approx 2.5 \times 10^7$. To understand how $g^{(2)}(0)$ is modulated by $\Omega$ and $\Omega_{wg}$, by Eq.~\eqref{eq:alpha}, we derive the steady-state $g^{(2)}$ at $\tau = 0$ for one transmon:

\begin{equation}
    g^{(2)}(0) \! = \! \frac{\left(\frac{\abs{\Omega}^2}{2} \! + \! \frac{\Gamma^2}{4}\right) \!\! \left[\frac{\abs{\Omega}^2}{2} \! + \! \frac{\Gamma^2}{4} + \Gamma^2 \!\!\left(\frac{4\abs{\Omega}^2}{\abs{\Omega_{wg}}^2} \! - \! 2Re[\frac{\Omega}{\Omega_{wg}}]\right)\right ]}{\left[\frac{\abs{\Omega}^2}{2} + \frac{\Gamma^2}{4} + \Gamma^2\left(\frac{\abs{\Omega}^2}{\abs{\Omega_{wg}}^2} - Re[\frac{\Omega}{\Omega_{wg}}]\right)\right]^2}.
    \label{eq:g2_theory}
\end{equation}
One can check that given a fixed $\Gamma$, $g^{(2)}(0)$ not only depends on the norm of the overall Rabi frequency $\Omega$, but is also determined by the norm and the real part of the ratio $\Omega / \Omega_{wg}$:\\
For case (1), $\Omega \ll \Gamma$ and the ratio $\Omega / \Omega_{wg} = 1$, rendering $g^{(2)}(0) \approx (\Gamma^2/4 + 2\Gamma^2) / (\Gamma^2/4) = 9$. \\
For case (2), one can take the limit $\Omega / \Omega_{wg} \rightarrow \infty$ so that $g^{(2)}(0) \rightarrow 1/(\abs{\Omega}^2 / \abs{\Omega_{wg}}^2) = 0$. \\
For case (3), $\Omega / \Omega_{wg} = 0.5$. As a result, $g^{(2)}(0) = [(\abs{\Omega}^2 / 2 + \Gamma^2 / 4) / (\abs{\Omega}^2/2)]^2 = 1 + \Gamma^2/ \abs{\Omega}^2 + \Gamma^4 / (4 \abs{\Omega}^4) \rightarrow \infty$ as $\Omega \rightarrow 0$.\\
Besides, one can obtain a coherent light with $g^{(2)}(0) = 1$ by letting $\Omega_d = -\Omega_{wg}$, rendering overall Rabi frequency $\Omega = 0$ at the transmon point. As a result, the transmon stays in the ground state and all the transmitted light is just the incident coherent light itself. The reason for the totally different behavior of the cases above lies in the different roles the two drives play: The waveguide driving $\Omega_{wg}$ not only manipulates the state of the transmon by Eq.~\eqref{eq:overall_rabi_freq} but also contributes to the interference in the transmitted field by Eq.~\eqref{eq:alpha}. The local drive, however, only appears in the Hamiltonian via Eq.~\eqref{eq:overall_rabi_freq}, with no accompanying photon in the waveguide. As a result, this local drive does not directly contribute to the interference in Eq.~\eqref{eq:alpha}. With the same overall Rabi frequency $\Omega$, case (1) $\sim$ (3) share the same steady state of the transmon, whose inchohernt component of emission should have the same power spectrum density. But the $g^{(2)}$ function takes the total transmitted field into account and the different field interference parameterized by $\Omega_{wg}$ yields very different photon statistics. 

In addition to the purity ($g^{(2)}$), one might also take the brightness into account while engineering the field. For example, in case (3), the terms $\Gamma^2/ \abs{\Omega}^2$ and $\Gamma^4 / (4 \abs{\Omega}^4)$ in the analytical expression of $g^{(2)}(0)$ indicate the purity/brightness trade-off because the photon bunching can be washed out by the large photon flux as $\Omega \rightarrow \infty$. However, the highly tunable photon statistics still hold at a brighter operating point. For example, at $\Omega = 0.1 \Gamma$, where the photon pair flux, $\langle \alpha^\dagger \alpha \rangle^2$ is $4$ orders of magnitude higher than $\Omega = 0.01 \Gamma$, $g^{(2)}(0) = 7.84, \ 0, \ 2.6 \times 10^3$ for case (1), (2), and (3), respectively.

\begin{figure}[!htbp]
  \centering
  \includegraphics[width = 0.5\textwidth]{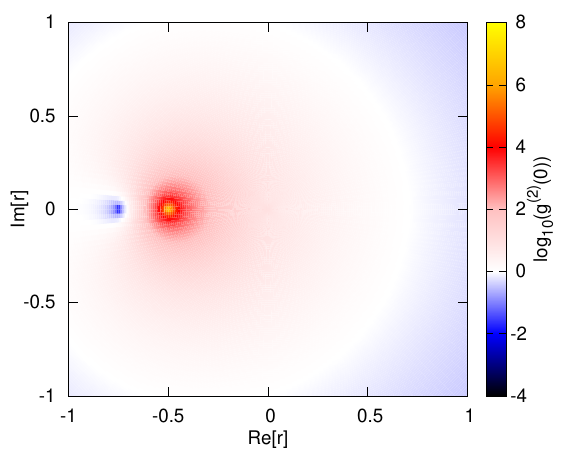}
  \caption{$g^{(2)}(0)$ as a function of the real part and imaginary part of $r$ while fixing $\Omega_{wg} = 0.02\Gamma$ and $\Delta = 0$.}
  \label{fig:g2_0_swp}
\end{figure}

In the cases above, the relative phase between $\Omega_{wg}$ and $\Omega_d$ is either $0$ or $\pi$. Without loss of generality, one can assume that $\Omega_{wg}$ is a real number and explore the amplitude and relative phase between $\Omega_{wg}$ and $\Omega_d$, parameterized by the ratio $r = \frac{\Omega_d}{\Omega_{wg}} = \abs{r}e^{i\phi}$, in the complex plane. By setting $\Omega_{wg} = 0.02\Gamma$ and $\Delta = 0$, Fig.~\ref{fig:g2_0_swp} shows $g^{(2)}(0)$ as a function of the real part and imaginary part of $r$.

One can notice the bright spot at (-0.5, 0), which corresponds to case (3), and the gradually decreased $g^{(2)}(0)$ as $\abs{r} \rightarrow \infty$, which asymptotically goes to case (2). The origin (0, 0) corresponds to case (1). In addition, there is a dip at (-0.75, 0), where both $\Omega_{wg}$ and $\Omega_d$ are nonzero. This suggests that the interference between the waveguide driving and the forward emission of the transmon can also give rise to an anti-bunched field. Other than the bright spot and the dip, the smooth modulation of $g^{(2)}(0)$ on the complex plane suggests that, by varying $\Omega_{wg}$ and $\Omega_d$, one can continuously manipulate the photon statistics of the transmitted field spanning anti-bunching, coherent, bunching, and super-bunching regimes.

\section{Summary\label{sec:summary}}
Using semi-classical and full-quantum calculations, we investigate the light after a chain of transmons chirally coupled to a one-dimensional waveguide. The semi-classical simulation results show that one chiral transmon can convert coherent light, with photons having long-range coherence and Poisson distribution, into a field with spatially localized coherence and bunching photon statistics. This is reminiscent of the nonlinear optics induced by the chiral transmon. Moreover, this process can be reversed - another chiral transmon applied downstream can completely restore the coherence of the light regardless of the Rabi frequency. This motivates us to investigate the processes yielding the incoherent light after one chiral transmon in the quantum regime.

We consider two-photon processes. With increasing width of the input wavepacket, $\Delta x$, the transmitted wavepacket exhibits power spectrum density of the incoherent component and $g^{(2)}$ function converging to the results from weak coherent drive. Together with the low photon flux of the weak coherent field, we infer that its incoherent transmission mainly comes from the two-photon interaction process. Moreover, the anti-correlation of $\abs{\Psi_{inc}^{sym}}^2$ in momentum space indicates the formation of a momentum-entangled photon pair from two independent incoming photons. This confirms the nonlinear quantum optics mediated by a single chiral transmon. The orthogonality of this entangled photon pair with the incoming wavepacket accounts for the loss of coherence after one chiral transmon. Meanwhile, the suppressed population of this photon pair induced by another chiral transmon explains the revival of coherence. In addition, the negativity in the Wigner function reveals the nonclassicality of the field. 

In addition to the coherent drive through the waveguide ($\Omega_{wg}$), we introduce a local drive with controllable amplitude and phase ($\Omega_d$) independent of the waveguide driving. For one chiral transmon, we show that by tuning $\Omega_d$ and $\Omega_{wg}$, one can control the contribution of the incoherent transmission and qualitatively alter the intensity correlation of the transmitted light. The analytical expression of $g^{(2)}(0)$ shows that the photon statistics can take the value from $0$ to $\infty$ by tuning $\Omega$ and the ratio $\Omega_{wg}/\Omega$, rendering the transmitted light anti-bunching, coherent, bunching, or super-bunching. With only one emitter and two kinds of coherent driving, this engineered driving allows a simple scheme to continuously modify the photon statistics of the output field \textit{in situ} via the amplitude and phase modulation of the local drive without changing the device and the waveguide input. Instead of the waveguide-only driving, this scheme allows the generation of the purely incoherent field, which is dominated by the momentum-correlated photon pairs, at weak driving. These photon pairs can be used for entanglement distribution in quantum networks, a minimal realization of few-photon nonlinear optics with a single emitter~\cite{PhysRevX.10.031011,Prasad2020}.

The main challenge of implementing the engineered driving lies in the local drive. In the superconducting qubit platform, the local drive can be accessible via a separate waveguide~\cite{Mirhosseini2019}, an individual driving line~\cite{Ma2019}, or ports in the sidewall of the waveguide~\cite{Zanner2022}. As for the observables in this work, the power spectrum density of the incoherent component can be detected using heterodyne detection~\cite{PhysRevX.13.021039,Mirhosseini2019}, while the photon correlation can be obtained by the moment-based correlation measurements~\cite{PhysRevA.82.043804,Bozyigit2011,PhysRevLett.106.243601,Lu2021}. Other candidate platforms include neutral atoms coupled to an optical nanofiber, with optical tweezers or separate addressing beams as the local drive~\cite{Kim2019,doi:10.1126/science.abi9917}, and quantum dots in semiconductors coupled to photonic-crystal waveguides~\cite{Sollner2015,Coles2016}, where the local drive can be implemented by the resonant excitation beam.

Although the formation of dimers~\cite{Stannigel_2012,PhysRevLett.113.237203,PhysRevA.91.042116} and Bethe Ansatz~\cite{PhysRevA.107.013717,PhysRevX.10.031011} can explain the coherent light after even numbers of spins chirally coupled to a one-dimensional waveguide, the detailed mechanism of restoring the coherence is still vague. Moreover, the full-quantum calculations presented in this work is limited to mimic the weak driving case. In the strongly driven case, the revival of coherence still holds even though the $N_{ph}$-photon processes with $N_{ph} > 2$ also substantially contribute to the generation of incoherent light. Further investigation of this regime calls for new theoretical methods. In addition, Ref.~\cite{PhysRevA.97.023813} gives an interference-based interpretation to the photon statistics of light scattered by one (two) qubit(s) bidirectionally coupled to a one-dimensional waveguide. Extending this method to the chiral case could provide a complementary perspective on the phenomenon observed in this work. In addition to only one transmon, the engineered driving can be also extended to the multi-transmon case, to see how the local drive can be used to control the properties of the transmitted field in the manybody systems.

\begin{acknowledgments}
The authors would like to thank Mohamed Eltohfa and AbdAlGhaffar Amer for useful discussions and suggestions.

This work was supported by the National Science Foundation under Award No. 2410890-PHY (WT, SW, and FR). Research supported as part of QuPIDC, an Energy Frontier Research Center, funded by the US Department of Energy (DOE), Office of Science, Basic Energy Sciences (BES), under award number DE-SC0025620 (HA). This research was supported in part through computational resources provided by Information Technology at Purdue University, West Lafayette, Indiana.
\end{acknowledgments}

\appendix

\section{Steady-state Power Spectrum Density and $g^{(2)}$ Function for Semi-classical Simulations\label{app:g1_g2}}
Although the photon part is traced out in the semi-classical calculations, the first-order ($g^{(1)}(\tau)$) and second-order ($g^{(2)}(\tau)$) correlation function of the steady-state transmitted field can be evaluated by the waveguide input-output theory and quantum regression theorem~\cite{loudon2000quantum, PhysRev.129.2342}. By definition, the first-order correlation function reads:
\begin{equation}
    g^{(1)}(\tau) = \frac{\langle \hat{\alpha}^\dagger (t) \hat{\alpha} (t + \tau) \rangle}{\langle \hat{\alpha}^\dagger (t) \hat{\alpha} (t) \rangle},
\end{equation}
and the power spectrum density is the Fourier transform of $g^{(1)}(\tau)$:
\begin{equation}
    G^{(1)}(\omega) = \frac{1}{2\pi} \int_{-\infty}^{+\infty} g^{(1)}(\tau) e^{i\omega \tau}d\tau.
    \label{eq:G1}
\end{equation}
The numerical calculation of $g^{(1)}(\tau)$ consists of three steps: First, we act $\hat{\alpha}^\dagger$ operator to the right of the steady-state density operator and normalize the result:
\begin{equation}
    \Tilde{\hat{\rho}}(0) = \frac{\hat{\rho}(\infty)\hat{\alpha}^\dagger}{\langle \hat{\alpha}^\dagger \hat{\alpha} \rangle},
\end{equation}
where $\langle \hat{\alpha}^\dagger \hat{\alpha} \rangle$ is evaluated in the steady state and is considered translational invariant. Second, we use $\Tilde{\hat{\rho}}(0)$ as the initial state and evolve $\Tilde{\hat{\rho}}(\tau)$ by the equations of motion in Sec.~\ref{sec:loss_and_revive_of_coherence}. Finally, the $g^{(1)}$ function of the total transmitted field can be evaluated by:
\begin{equation}
    g^{(1)}_{tot}(\tau) = Tr(\hat{\alpha}\Tilde{\hat{\rho}}(\tau)).
\end{equation}
This $g^{(1)}$ function contains both the components of the transmission that can/cannot interfere with the incident light and one can check that $g^{(1)}_{tot}(\tau \rightarrow \infty) = T_{coh}/T_{tot}$, which will contribute to a spike in the power spectrum density. To obtain the power spectrum of the component that cannot interfere, we remove the asymptotic value of $g^{(1)}$:
\begin{equation}
    g^{(1)}_{inc}(\tau) = g^{(1)}_{tot}(\tau) - \frac{T_{coh}}{T_{tot}}
\end{equation}
Note that the numerical calculation only yields the $g^{(1)}(\tau)$ when $\tau \ge 0$. For $\tau < 0$, one can utilize the fact that:
\begin{equation}
    g^{(1)}(-\tau) = g^{(1)*}(\tau).
\end{equation}

The $g^{(2)}$ function, defined by:
\begin{equation}
    g^{(2)}(\tau) = \frac{\langle \hat{\alpha}^\dagger (t) \hat{\alpha}^\dagger (t + \tau) \hat{\alpha} (t + \tau) \hat{\alpha} (t) \rangle}{\langle \hat{\alpha}^\dagger (t) \hat{\alpha} (t) \rangle \langle \hat{\alpha}^\dagger (t + \tau) \hat{\alpha} (t + \tau) \rangle},
\end{equation}
can be evaluated similarly to $g^{(1)}$, except that $\Tilde{\hat{\rho}}(0)$ is obtained by:
\begin{equation}
    \Tilde{\hat{\rho}}(0) = \frac{\hat{\alpha} \hat{\rho}(\infty) \hat{\alpha}^\dagger}{\langle \hat{\alpha}^\dagger \hat{\alpha} \rangle},
\end{equation}
and $g^{(2)}(\tau)$ is evaluated by:
\begin{equation}
    g^{(2)}(\tau) =  \frac{Tr(\hat{\alpha} \Tilde{\hat{\rho}}(\tau) \hat{\alpha}^\dagger)}{\langle \hat{\alpha}^\dagger \hat{\alpha} \rangle}.
\end{equation}
\newpage

\bibliography{ref.bib}

@Article{Lodahl2017,
author={Lodahl, Peter
and Mahmoodian, Sahand
and Stobbe, S{\o}ren
and Rauschenbeutel, Arno
and Schneeweiss, Philipp
and Volz, J{\"u}rgen
and Pichler, Hannes
and Zoller, Peter},
title={Chiral quantum optics},
journal={Nature},
year={2017},
month={Jan},
day={01},
volume={541},
number={7638},
pages={473-480},
issn={1476-4687},
doi={10.1038/nature21037},
url={https://doi.org/10.1038/nature21037}
}

@article{PRXQuantum.6.020101,
  title = {Chiral Quantum Optics: Recent Developments and Future Directions},
  author = {Su\'arez-Forero, D.G. and Jalali Mehrabad, M. and Vega, C. and Gonz\'alez-Tudela, A. and Hafezi, M.},
  journal = {PRX Quantum},
  volume = {6},
  issue = {2},
  pages = {020101},
  numpages = {18},
  year = {2025},
  month = {Apr},
  publisher = {American Physical Society},
  doi = {10.1103/PRXQuantum.6.020101},
  url = {https://link.aps.org/doi/10.1103/PRXQuantum.6.020101}
}

@article{PhysRevLett.78.3221,
  title = {Quantum State Transfer and Entanglement Distribution among Distant Nodes in a Quantum Network},
  author = {Cirac, J. I. and Zoller, P. and Kimble, H. J. and Mabuchi, H.},
  journal = {Phys. Rev. Lett.},
  volume = {78},
  issue = {16},
  pages = {3221--3224},
  numpages = {0},
  year = {1997},
  month = {Apr},
  publisher = {American Physical Society},
  doi = {10.1103/PhysRevLett.78.3221},
  url = {https://link.aps.org/doi/10.1103/PhysRevLett.78.3221}
}

@Article{Kurpiers2018,
author={Kurpiers, P.
and Magnard, P.
and Walter, T.
and Royer, B.
and Pechal, M.
and Heinsoo, J.
and Salath{\'e}, Y.
and Akin, A.
and Storz, S.
and Besse, J.-C.
and Gasparinetti, S.
and Blais, A.
and Wallraff, A.},
title={Deterministic quantum state transfer and remote entanglement using microwave photons},
journal={Nature},
year={2018},
month={Jun},
day={01},
volume={558},
number={7709},
pages={264-267},
issn={1476-4687},
doi={10.1038/s41586-018-0195-y},
url={https://doi.org/10.1038/s41586-018-0195-y}
}

@Article{Northup2014,
author={Northup, T. E.
and Blatt, R.},
title={Quantum information transfer using photons},
journal={Nature Photonics},
year={2014},
month={May},
day={01},
volume={8},
number={5},
pages={356-363},
doi={10.1038/nphoton.2014.53},
url={https://doi.org/10.1038/nphoton.2014.53}
}

@Article{Axline2018,
author={Axline, Christopher J.
and Burkhart, Luke D.
and Pfaff, Wolfgang
and Zhang, Mengzhen
and Chou, Kevin
and Campagne-Ibarcq, Philippe
and Reinhold, Philip
and Frunzio, Luigi
and Girvin, S. M.
and Jiang, Liang
and Devoret, M. H.
and Schoelkopf, R. J.},
title={On-demand quantum state transfer and entanglement between remote microwave cavity memories},
journal={Nature Physics},
year={2018},
month={Jul},
day={01},
volume={14},
number={7},
pages={705-710},
issn={1745-2481},
doi={10.1038/s41567-018-0115-y},
url={https://doi.org/10.1038/s41567-018-0115-y}
}

@article{PhysRevLett.120.200501,
  title = {Deterministic Remote Entanglement of Superconducting Circuits through Microwave Two-Photon Transitions},
  author = {Campagne-Ibarcq, P. and Zalys-Geller, E. and Narla, A. and Shankar, S. and Reinhold, P. and Burkhart, L. and Axline, C. and Pfaff, W. and Frunzio, L. and Schoelkopf, R. J. and Devoret, M. H.},
  journal = {Phys. Rev. Lett.},
  volume = {120},
  issue = {20},
  pages = {200501},
  numpages = {6},
  year = {2018},
  month = {May},
  publisher = {American Physical Society},
  doi = {10.1103/PhysRevLett.120.200501},
  url = {https://link.aps.org/doi/10.1103/PhysRevLett.120.200501}
}

@article{PhysRevApplied.12.044067,
  title = {Quantum Communication with Time-Bin Encoded Microwave Photons},
  author = {Kurpiers, P. and Pechal, M. and Royer, B. and Magnard, P. and Walter, T. and Heinsoo, J. and Salath\'e, Y. and Akin, A. and Storz, S. and Besse, J.-C. and Gasparinetti, S. and Blais, A. and Wallraff, A.},
  journal = {Phys. Rev. Appl.},
  volume = {12},
  issue = {4},
  pages = {044067},
  numpages = {11},
  year = {2019},
  month = {Oct},
  publisher = {American Physical Society},
  doi = {10.1103/PhysRevApplied.12.044067},
  url = {https://link.aps.org/doi/10.1103/PhysRevApplied.12.044067}
}

@article{PhysRevLett.125.260502,
  title = {Microwave Quantum Link between Superconducting Circuits Housed in Spatially Separated Cryogenic Systems},
  author = {Magnard, P. and Storz, S. and Kurpiers, P. and Sch\"ar, J. and Marxer, F. and L\"utolf, J. and Walter, T. and Besse, J.-C. and Gabureac, M. and Reuer, K. and Akin, A. and Royer, B. and Blais, A. and Wallraff, A.},
  journal = {Phys. Rev. Lett.},
  volume = {125},
  issue = {26},
  pages = {260502},
  numpages = {7},
  year = {2020},
  month = {Dec},
  publisher = {American Physical Society},
  doi = {10.1103/PhysRevLett.125.260502},
  url = {https://link.aps.org/doi/10.1103/PhysRevLett.125.260502}
}

@article{PhysRevLett.118.133601,
  title = {Quantum State Transfer via Noisy Photonic and Phononic Waveguides},
  author = {Vermersch, B. and Guimond, P.-O. and Pichler, H. and Zoller, P.},
  journal = {Phys. Rev. Lett.},
  volume = {118},
  issue = {13},
  pages = {133601},
  numpages = {6},
  year = {2017},
  month = {Mar},
  publisher = {American Physical Society},
  doi = {10.1103/PhysRevLett.118.133601},
  url = {https://link.aps.org/doi/10.1103/PhysRevLett.118.133601}
}

@article{PhysRevX.7.011035,
  title = {Intracity Quantum Communication via Thermal Microwave Networks},
  author = {Xiang, Ze-Liang and Zhang, Mengzhen and Jiang, Liang and Rabl, Peter},
  journal = {Phys. Rev. X},
  volume = {7},
  issue = {1},
  pages = {011035},
  numpages = {11},
  year = {2017},
  month = {Mar},
  publisher = {American Physical Society},
  doi = {10.1103/PhysRevX.7.011035},
  url = {https://link.aps.org/doi/10.1103/PhysRevX.7.011035}
}

@article{PhysRevResearch.2.013369,
  title = {Long-distance dissipation-assisted transport of entangled states via a chiral waveguide},
  author = {Mok, Wai-Keong and Aghamalyan, Davit and You, Jia-Bin and Haug, Tobias and Zhang, Wenzu and Png, Ching Eng and Kwek, Leong-Chuan},
  journal = {Phys. Rev. Res.},
  volume = {2},
  issue = {1},
  pages = {013369},
  numpages = {7},
  year = {2020},
  month = {Mar},
  publisher = {American Physical Society},
  doi = {10.1103/PhysRevResearch.2.013369},
  url = {https://link.aps.org/doi/10.1103/PhysRevResearch.2.013369}
}

@article{Yu_2026,
doi = {10.1088/1367-2630/ae481f},
url = {https://doi.org/10.1088/1367-2630/ae481f},
year = {2026},
month = {feb},
publisher = {IOP Publishing},
volume = {28},
number = {3},
pages = {034501},
author = {Yu, Ya-Tang and Handayana, I Gusti Ngurah Yudi and Chen, Wei and Jen, H H},
title = {Fast and high excitation transport in waveguide quantum electrodynamics},
journal = {New Journal of Physics}
}

@article{PhysRevA.103.063711,
  title = {Bound and subradiant multiatom excitations in an atomic array with nonreciprocal couplings},
  author = {Jen, H. H.},
  journal = {Phys. Rev. A},
  volume = {103},
  issue = {6},
  pages = {063711},
  numpages = {7},
  year = {2021},
  month = {Jun},
  publisher = {American Physical Society},
  doi = {10.1103/PhysRevA.103.063711},
  url = {https://link.aps.org/doi/10.1103/PhysRevA.103.063711}
}

@article{Jen_2019,
doi = {10.1088/1361-6455/ab04c1},
url = {https://doi.org/10.1088/1361-6455/ab04c1},
year = {2019},
month = {mar},
publisher = {IOP Publishing},
volume = {52},
number = {6},
pages = {065502},
author = {Jen, H H},
title = {Selective transport of atomic excitations in a driven chiral-coupled atomic chain},
journal = {Journal of Physics B: Atomic, Molecular and Optical Physics},
}

@article{Stannigel_2012,
doi = {10.1088/1367-2630/14/6/063014},
url = {https://doi.org/10.1088/1367-2630/14/6/063014},
year = {2012},
month = {jun},
publisher = {IOP Publishing},
volume = {14},
number = {6},
pages = {063014},
author = {Stannigel, K and Rabl, P and Zoller, P},
title = {Driven-dissipative preparation of entangled states in cascaded quantum-optical networks},
journal = {New Journal of Physics},
}

@article{PhysRevLett.113.237203,
  title = {Quantum Spin Dimers from Chiral Dissipation in Cold-Atom Chains},
  author = {Ramos, Tom\'as and Pichler, Hannes and Daley, Andrew J. and Zoller, Peter},
  journal = {Phys. Rev. Lett.},
  volume = {113},
  issue = {23},
  pages = {237203},
  numpages = {6},
  year = {2014},
  month = {Dec},
  publisher = {American Physical Society},
  doi = {10.1103/PhysRevLett.113.237203},
  url = {https://link.aps.org/doi/10.1103/PhysRevLett.113.237203}
}

@article{PhysRevA.91.042116,
  title = {Quantum optics of chiral spin networks},
  author = {Pichler, Hannes and Ramos, Tom\'as and Daley, Andrew J. and Zoller, Peter},
  journal = {Phys. Rev. A},
  volume = {91},
  issue = {4},
  pages = {042116},
  numpages = {19},
  year = {2015},
  month = {Apr},
  publisher = {American Physical Society},
  doi = {10.1103/PhysRevA.91.042116},
  url = {https://link.aps.org/doi/10.1103/PhysRevA.91.042116}
}

@article{BLIOKH20151,
title = {Transverse and longitudinal angular momenta of light},
journal = {Physics Reports},
volume = {592},
pages = {1-38},
year = {2015},
note = {Transverse and longitudinal angular momenta of light},
issn = {0370-1573},
doi = {https://doi.org/10.1016/j.physrep.2015.06.003},
url = {https://www.sciencedirect.com/science/article/pii/S0370157315003336},
author = {Konstantin Y. Bliokh and Franco Nori},
}

@Article{Bliokh2015,
author={Bliokh, K. Y.
and Rodr{\'i}guez-Fortu{\~{n}}o, F. J.
and Nori, F.
and Zayats, A. V.},
title={Spin--orbit interactions of light},
journal={Nature Photonics},
year={2015},
month={Dec},
day={01},
volume={9},
number={12},
pages={796-808},
issn={1749-4893},
doi={10.1038/nphoton.2015.201},
url={https://doi.org/10.1038/nphoton.2015.201}
}

@Article{Aiello2015,
author={Aiello, Andrea
and Banzer, Peter
and Neugebauer, Martin
and Leuchs, Gerd},
title={From transverse angular momentum to photonic wheels},
journal={Nature Photonics},
year={2015},
month={Dec},
day={01},
volume={9},
number={12},
pages={789-795},
issn={1749-4893},
doi={10.1038/nphoton.2015.203},
url={https://doi.org/10.1038/nphoton.2015.203}
}

@article{Sinev2017,
author = {Sinev, Ivan S. and Bogdanov, Andrey A. and Komissarenko, Filipp E. and Frizyuk, Kristina S. and Petrov, Mihail I. and Mukhin, Ivan S. and Makarov, Sergey V. and Samusev, Anton K. and Lavrinenko, Andrei V. and Iorsh, Ivan V.},
title = {Chirality Driven by Magnetic Dipole Response for Demultiplexing of Surface Waves},
journal = {Laser \& Photonics Reviews},
volume = {11},
number = {5},
pages = {1700168},
doi = {https://doi.org/10.1002/lpor.201700168},
url = {https://onlinelibrary.wiley.com/doi/abs/10.1002/lpor.201700168},
year = {2017}
}

@article{Francisco2013,
author = {Francisco J. Rodríguez-Fortuño  and Giuseppe Marino  and Pavel Ginzburg  and Daniel O’Connor  and Alejandro Martínez  and Gregory A. Wurtz  and Anatoly V. Zayats },
title = {Near-Field Interference for the Unidirectional Excitation of Electromagnetic Guided Modes},
journal = {Science},
volume = {340},
number = {6130},
pages = {328-330},
year = {2013},
doi = {10.1126/science.1233739},
URL = {https://www.science.org/doi/abs/10.1126/science.1233739},
}

@Article{Spitzer2018,
author={Spitzer, F.
and Poddubny, A. N.
and Akimov, I. A.
and Sapega, V. F.
and Klompmaker, L.
and Kreilkamp, L. E.
and Litvin, L. V.
and Jede, R.
and Karczewski, G.
and Wiater, M.
and Wojtowicz, T.
and Yakovlev, D. R.
and Bayer, M.},
title={Routing the emission of a near-surface light source by a magnetic field},
journal={Nature Physics},
year={2018},
month={Oct},
day={01},
volume={14},
number={10},
pages={1043-1048},
issn={1745-2481},
doi={10.1038/s41567-018-0232-7},
url={https://doi.org/10.1038/s41567-018-0232-7}
}

@article{PhysRevLett.113.243601,
  title = {Fabry-Perot Interferometer with Quantum Mirrors: Nonlinear Light Transport and Rectification},
  author = {Fratini, F. and Mascarenhas, E. and Safari, L. and Poizat, J-Ph. and Valente, D. and Auff\`eves, A. and Gerace, D. and Santos, M. F.},
  journal = {Phys. Rev. Lett.},
  volume = {113},
  issue = {24},
  pages = {243601},
  numpages = {5},
  year = {2014},
  month = {Dec},
  publisher = {American Physical Society},
  doi = {10.1103/PhysRevLett.113.243601},
  url = {https://link.aps.org/doi/10.1103/PhysRevLett.113.243601}
}

@article{PhysRevLett.121.123601,
  title = {Nonreciprocity Realized with Quantum Nonlinearity},
  author = {Rosario Hamann, Andr\'es and M\"uller, Clemens and Jerger, Markus and Zanner, Maximilian and Combes, Joshua and Pletyukhov, Mikhail and Weides, Martin and Stace, Thomas M. and Fedorov, Arkady},
  journal = {Phys. Rev. Lett.},
  volume = {121},
  issue = {12},
  pages = {123601},
  numpages = {5},
  year = {2018},
  month = {Sep},
  publisher = {American Physical Society},
  doi = {10.1103/PhysRevLett.121.123601},
  url = {https://link.aps.org/doi/10.1103/PhysRevLett.121.123601}
}

@article{PhysRevA.102.053720,
  title = {Programmable directional emitter and receiver of itinerant microwave photons in a waveguide},
  author = {Gheeraert, Nicolas and Kono, Shingo and Nakamura, Yasunobu},
  journal = {Phys. Rev. A},
  volume = {102},
  issue = {5},
  pages = {053720},
  numpages = {14},
  year = {2020},
  month = {Nov},
  publisher = {American Physical Society},
  doi = {10.1103/PhysRevA.102.053720},
  url = {https://link.aps.org/doi/10.1103/PhysRevA.102.053720}
}

@Article{Guimond2020,
author={Guimond, P.-O.
and Vermersch, B.
and Juan, M. L.
and Sharafiev, A.
and Kirchmair, G.
and Zoller, P.},
title={A unidirectional on-chip photonic interface for superconducting circuits},
journal={npj Quantum Information},
year={2020},
month={Mar},
day={27},
volume={6},
number={1},
pages={32},
issn={2056-6387},
doi={10.1038/s41534-020-0261-9},
url={https://doi.org/10.1038/s41534-020-0261-9}
}

@Article{Kannan2023,
author={Kannan, Bharath
and Almanakly, Aziza
and Sung, Youngkyu
and Di Paolo, Agustin
and Rower, David A.
and Braum{\"u}ller, Jochen
and Melville, Alexander
and Niedzielski, Bethany M.
and Karamlou, Amir
and Serniak, Kyle
and Veps{\"a}l{\"a}inen, Antti
and Schwartz, Mollie E.
and Yoder, Jonilyn L.
and Winik, Roni
and Wang, Joel I-Jan
and Orlando, Terry P.
and Gustavsson, Simon
and Grover, Jeffrey A.
and Oliver, William D.},
title={On-demand directional microwave photon emission using waveguide quantum electrodynamics},
journal={Nature Physics},
year={2023},
month={Mar},
day={01},
volume={19},
number={3},
pages={394-400},
issn={1745-2481},
doi={10.1038/s41567-022-01869-5},
url={https://doi.org/10.1038/s41567-022-01869-5}
}

@Article{Redchenko2023,
author={Redchenko, Elena S.
and Poshakinskiy, Alexander V.
and Sett, Riya
and {\v{Z}}emli{\v{c}}ka, Martin
and Poddubny, Alexander N.
and Fink, Johannes M.},
title={Tunable directional photon scattering from a pair of superconducting qubits},
journal={Nature Communications},
year={2023},
month={May},
day={24},
volume={14},
number={1},
pages={2998},
doi={10.1038/s41467-023-38761-6},
url={https://doi.org/10.1038/s41467-023-38761-6}
}

@article{PhysRevX.13.021039,
  title = {Resonance Fluorescence of a Chiral Artificial Atom},
  author = {Joshi, Chaitali and Yang, Frank and Mirhosseini, Mohammad},
  journal = {Phys. Rev. X},
  volume = {13},
  issue = {2},
  pages = {021039},
  numpages = {27},
  year = {2023},
  month = {Jun},
  publisher = {American Physical Society},
  doi = {10.1103/PhysRevX.13.021039},
  url = {https://link.aps.org/doi/10.1103/PhysRevX.13.021039}
}

@article{PhysRevLett.110.213604,
  title = {Strong Coupling between Single Atoms and Nontransversal Photons},
  author = {Junge, Christian and O'Shea, Danny and Volz, J\"urgen and Rauschenbeutel, Arno},
  journal = {Phys. Rev. Lett.},
  volume = {110},
  issue = {21},
  pages = {213604},
  numpages = {5},
  year = {2013},
  month = {May},
  publisher = {American Physical Society},
  doi = {10.1103/PhysRevLett.110.213604},
  url = {https://link.aps.org/doi/10.1103/PhysRevLett.110.213604}
}

@article{Itay2014,
author = {Itay Shomroni  and Serge Rosenblum  and Yulia Lovsky  and Orel Bechler  and Gabriel Guendelman  and Barak Dayan },
title = {All-optical routing of single photons by a one-atom switch controlled by a single photon},
journal = {Science},
volume = {345},
number = {6199},
pages = {903-906},
year = {2014},
doi = {10.1126/science.1254699},
URL = {https://www.science.org/doi/abs/10.1126/science.1254699},
}

@Article{Mitsch2014,
author={Mitsch, R.
and Sayrin, C.
and Albrecht, B.
and Schneeweiss, P.
and Rauschenbeutel, A.},
title={Quantum state-controlled directional spontaneous emission of photons into a nanophotonic waveguide},
journal={Nature Communications},
year={2014},
month={Dec},
day={12},
volume={5},
number={1},
pages={5713},
issn={2041-1723},
doi={10.1038/ncomms6713},
url={https://doi.org/10.1038/ncomms6713}
}

@article{PhysRevLett.110.037402,
  title = {Interfacing Spins in an InGaAs Quantum Dot to a Semiconductor Waveguide Circuit Using Emitted Photons},
  author = {Luxmoore, I. J. and Wasley, N. A. and Ramsay, A. J. and Thijssen, A. C. T. and Oulton, R. and Hugues, M. and Kasture, S. and Achanta, V. G. and Fox, A. M. and Skolnick, M. S.},
  journal = {Phys. Rev. Lett.},
  volume = {110},
  issue = {3},
  pages = {037402},
  numpages = {5},
  year = {2013},
  month = {Jan},
  publisher = {American Physical Society},
  doi = {10.1103/PhysRevLett.110.037402},
  url = {https://link.aps.org/doi/10.1103/PhysRevLett.110.037402}
}

@article{Luxmoore2013,
    author = {Luxmoore, I. J. and Wasley, N. A. and Ramsay, A. J. and Thijssen, A. C. T. and Oulton, R. and Hugues, M. and Fox, A. M. and Skolnick, M. S.},
    title = {Optical control of the emission direction of a quantum dot},
    journal = {Applied Physics Letters},
    volume = {103},
    number = {24},
    pages = {241102},
    year = {2013},
    month = {12},
    issn = {0003-6951},
    doi = {10.1063/1.4845975},
    url = {https://doi.org/10.1063/1.4845975},
}

@article{Sollner2015,
author={S{\"o}llner, Immo
and Mahmoodian, Sahand
and Hansen, Sofie Lindskov
and Midolo, Leonardo
and Javadi, Alisa
and Kir{\v{s}}ansk{\.{e}}, Gabija
and Pregnolato, Tommaso
and El-Ella, Haitham
and Lee, Eun Hye
and Song, Jin Dong
and Stobbe, S{\o}ren
and Lodahl, Peter},
title={Deterministic photon--emitter coupling in chiral photonic circuits},
journal={Nature Nanotechnology},
year={2015},
month={Sep},
day={01},
volume={10},
number={9},
pages={775-778},
issn={1748-3395},
doi={10.1038/nnano.2015.159},
url={https://doi.org/10.1038/nnano.2015.159}
}

@Article{OConnor2014,
author={O'Connor, D.
and Ginzburg, P.
and Rodr{\'i}guez-Fortu{\~{n}}o, F. J.
and Wurtz, G. A.
and Zayats, A. V.},
title={Spin--orbit coupling in surface plasmon scattering by nanostructures},
journal={Nature Communications},
year={2014},
month={Nov},
day={13},
volume={5},
number={1},
pages={5327},
issn={2041-1723},
doi={10.1038/ncomms6327},
url={https://doi.org/10.1038/ncomms6327}
}

@Article{Coles2016,
author={Coles, R. J.
and Price, D. M.
and Dixon, J. E.
and Royall, B.
and Clarke, E.
and Kok, P.
and Skolnick, M. S.
and Fox, A. M.
and Makhonin, M. N.},
title={Chirality of nanophotonic waveguide with embedded quantum emitter for unidirectional spin transfer},
journal={Nature Communications},
year={2016},
month={Mar},
day={31},
volume={7},
number={1},
pages={11183},
issn={2041-1723},
doi={10.1038/ncomms11183},
url={https://doi.org/10.1038/ncomms11183}
}

@Article{Neugebauer2014,
author={Neugebauer, Martin
and Bauer, Thomas
and Banzer, Peter
and Leuchs, Gerd},
title={Polarization Tailored Light Driven Directional Optical Nanobeacon},
journal={Nano Letters},
year={2014},
month={May},
day={14},
publisher={American Chemical Society},
volume={14},
number={5},
pages={2546-2551},
issn={1530-6984},
doi={10.1021/nl5003526},
url={https://doi.org/10.1021/nl5003526}
}

@article{Jan2014,
author = {Jan Petersen  and Jürgen Volz  and Arno Rauschenbeutel },
title = {Chiral nanophotonic waveguide interface based on spin-orbit interaction of light},
journal = {Science},
volume = {346},
number = {6205},
pages = {67-71},
year = {2014},
doi = {10.1126/science.1257671},
URL = {https://www.science.org/doi/abs/10.1126/science.1257671},
}

@article{PhysRevLett.108.213907,
  title = {Role of Magnetic Induction Currents in Nanoslit Excitation of Surface Plasmon Polaritons},
  author = {Lee, Seung-Yeol and Lee, Il-Min and Park, Junghyun and Oh, Sewoong and Lee, Wooyoung and Kim, Kyoung-Youm and Lee, Byoungho},
  journal = {Phys. Rev. Lett.},
  volume = {108},
  issue = {21},
  pages = {213907},
  numpages = {5},
  year = {2012},
  month = {May},
  publisher = {American Physical Society},
  doi = {10.1103/PhysRevLett.108.213907},
  url = {https://link.aps.org/doi/10.1103/PhysRevLett.108.213907}
}

@article{Jiao2013,
author = {Jiao Lin  and J. P. Balthasar Mueller  and Qian Wang  and Guanghui Yuan  and Nicholas Antoniou  and Xiao-Cong Yuan  and Federico Capasso },
title = {Polarization-Controlled Tunable Directional Coupling of Surface Plasmon Polaritons},
journal = {Science},
volume = {340},
number = {6130},
pages = {331-334},
year = {2013},
doi = {10.1126/science.1233746},
URL = {https://www.science.org/doi/abs/10.1126/science.1233746},}

@ARTICLE{Roth2023,
  author={Roth, Thomas E. and Ma, Ruichao and Chew, Weng C.},
  journal={IEEE Antennas and Propagation Magazine}, 
  title={The Transmon Qubit for Electromagnetics Engineers: An introduction}, 
  year={2023},
  volume={65},
  number={2},
  pages={8-20},
  doi={10.1109/MAP.2022.3176593}}

@article{PhysRevA.69.062320,
  title = {Cavity quantum electrodynamics for superconducting electrical circuits: An architecture for quantum computation},
  author = {Blais, Alexandre and Huang, Ren-Shou and Wallraff, Andreas and Girvin, S. M. and Schoelkopf, R. J.},
  journal = {Phys. Rev. A},
  volume = {69},
  issue = {6},
  pages = {062320},
  numpages = {14},
  year = {2004},
  month = {Jun},
  publisher = {American Physical Society},
  doi = {10.1103/PhysRevA.69.062320},
  url = {https://link.aps.org/doi/10.1103/PhysRevA.69.062320}
}

@article{PhysRevA.75.032329,
  title = {Quantum-information processing with circuit quantum electrodynamics},
  author = {Blais, Alexandre and Gambetta, Jay and Wallraff, A. and Schuster, D. I. and Girvin, S. M. and Devoret, M. H. and Schoelkopf, R. J.},
  journal = {Phys. Rev. A},
  volume = {75},
  issue = {3},
  pages = {032329},
  numpages = {21},
  year = {2007},
  month = {Mar},
  publisher = {American Physical Society},
  doi = {10.1103/PhysRevA.75.032329},
  url = {https://link.aps.org/doi/10.1103/PhysRevA.75.032329}
}

@article{GU20171,
title = {Microwave photonics with superconducting quantum circuits},
journal = {Physics Reports},
volume = {718-719},
pages = {1-102},
year = {2017},
note = {Microwave photonics with superconducting quantum circuits},
issn = {0370-1573},
doi = {https://doi.org/10.1016/j.physrep.2017.10.002},
url = {https://www.sciencedirect.com/science/article/pii/S0370157317303290},
author = {Xiu Gu and Anton Frisk Kockum and Adam Miranowicz and Yu-xi Liu and Franco Nori},
}

@article{Krantz2019,
    author = {Krantz, P. and Kjaergaard, M. and Yan, F. and Orlando, T. P. and Gustavsson, S. and Oliver, W. D.},
    title = {A quantum engineer's guide to superconducting qubits},
    journal = {Applied Physics Reviews},
    volume = {6},
    number = {2},
    pages = {021318},
    year = {2019},
    month = {06},
    issn = {1931-9401},
    doi = {10.1063/1.5089550},
    url = {https://doi.org/10.1063/1.5089550},
}

@Article{Wallraff2004,
author={Wallraff, A.
and Schuster, D. I.
and Blais, A.
and Frunzio, L.
and Huang, R.-. S.
and Majer, J.
and Kumar, S.
and Girvin, S. M.
and Schoelkopf, R. J.},
title={Strong coupling of a single photon to a superconducting qubit using circuit quantum electrodynamics},
journal={Nature},
year={2004},
month={Sep},
day={01},
volume={431},
number={7005},
pages={162-167},
issn={1476-4687},
doi={10.1038/nature02851},
url={https://doi.org/10.1038/nature02851}
}

@article{Arjan2013,
author = {Arjan F. van Loo  and Arkady Fedorov  and Kevin Lalumière  and Barry C. Sanders  and Alexandre Blais  and Andreas Wallraff },
title = {Photon-Mediated Interactions Between Distant Artificial Atoms},
journal = {Science},
volume = {342},
number = {6165},
pages = {1494-1496},
year = {2013},
doi = {10.1126/science.1244324},
URL = {https://www.science.org/doi/abs/10.1126/science.1244324},}

@Article{Ma2019,
author={Ma, Ruichao
and Saxberg, Brendan
and Owens, Clai
and Leung, Nelson
and Lu, Yao
and Simon, Jonathan
and Schuster, David I.},
title={A dissipatively stabilized Mott insulator of photons},
journal={Nature},
year={2019},
month={Feb},
day={01},
volume={566},
number={7742},
pages={51-57},
issn={1476-4687},
doi={10.1038/s41586-019-0897-9},
url={https://doi.org/10.1038/s41586-019-0897-9}
}

@misc{Botao2026,
      title={Programmable Superradiance in an Interacting Qubit Array}, 
      author={Botao Du and Qihao Guo and Ruichao Ma},
      year={2026},
      archivePrefix={arXiv},
      primaryClass={cond-mat.quant-gas},
      url={https://arxiv.org/abs/2605.12442}, 
}

@Article{Almanakly2025,
author={Almanakly, Aziza
and Yankelevich, Beatriz
and Hays, Max
and Kannan, Bharath
and Assouly, R{\'e}ouven
and Greene, Alex
and Gingras, Michael
and Niedzielski, Bethany M.
and Stickler, Hannah
and Schwartz, Mollie E.
and Serniak, Kyle
and Wang, Joel {\^I}-j.
and Orlando, Terry P.
and Gustavsson, Simon
and Grover, Jeffrey A.
and Oliver, William D.},
title={Deterministic remote entanglement using a chiral quantum interconnect},
journal={Nature Physics},
year={2025},
month={May},
day={01},
volume={21},
number={5},
pages={825-830},
issn={1745-2481},
doi={10.1038/s41567-025-02811-1},
url={https://doi.org/10.1038/s41567-025-02811-1}
}

@article{RevModPhys.95.015002,
  title = {Waveguide quantum electrodynamics: Collective radiance and photon-photon correlations},
  author = {Sheremet, Alexandra S. and Petrov, Mihail I. and Iorsh, Ivan V. and Poshakinskiy, Alexander V. and Poddubny, Alexander N.},
  journal = {Rev. Mod. Phys.},
  volume = {95},
  issue = {1},
  pages = {015002},
  numpages = {59},
  year = {2023},
  month = {Mar},
  publisher = {American Physical Society},
  doi = {10.1103/RevModPhys.95.015002},
  url = {https://link.aps.org/doi/10.1103/RevModPhys.95.015002}
}

@Article{Mirhosseini2019,
author={Mirhosseini, Mohammad
and Kim, Eunjong
and Zhang, Xueyue
and Sipahigil, Alp
and Dieterle, Paul B.
and Keller, Andrew J.
and Asenjo-Garcia, Ana
and Chang, Darrick E.
and Painter, Oskar},
title={Cavity quantum electrodynamics with atom-like mirrors},
journal={Nature},
year={2019},
month={May},
day={01},
volume={569},
number={7758},
pages={692-697},
issn={1476-4687},
doi={10.1038/s41586-019-1196-1},
url={https://doi.org/10.1038/s41586-019-1196-1}
}

@book{loudon2000quantum,
  title={The quantum theory of light},
  author={Loudon, Rodney},
  year={2000},
  publisher={OUP Oxford}
}

@article{PhysRev.129.2342,
  title = {Formal Theory of Quantum Fluctuations from a Driven State},
  author = {Lax, Melvin},
  journal = {Phys. Rev.},
  volume = {129},
  issue = {5},
  pages = {2342--2348},
  numpages = {0},
  year = {1963},
  month = {Mar},
  publisher = {American Physical Society},
  doi = {10.1103/PhysRev.129.2342},
  url = {https://link.aps.org/doi/10.1103/PhysRev.129.2342}
}

@article{PhysRevA.88.043806,
  title = {Input-output theory for waveguide QED with an ensemble of inhomogeneous atoms},
  author = {Lalumi\`ere, Kevin and Sanders, Barry C. and van Loo, A. F. and Fedorov, A. and Wallraff, A. and Blais, A.},
  journal = {Phys. Rev. A},
  volume = {88},
  issue = {4},
  pages = {043806},
  numpages = {15},
  year = {2013},
  month = {Oct},
  publisher = {American Physical Society},
  doi = {10.1103/PhysRevA.88.043806},
  url = {https://link.aps.org/doi/10.1103/PhysRevA.88.043806}
}

@article{PhysRevLett.95.213001,
  title = {Coherent Single Photon Transport in a One-Dimensional Waveguide Coupled with Superconducting Quantum Bits},
  author = {Shen, Jung-Tsung and Fan, Shanhui},
  journal = {Phys. Rev. Lett.},
  volume = {95},
  issue = {21},
  pages = {213001},
  numpages = {4},
  year = {2005},
  month = {Nov},
  publisher = {American Physical Society},
  doi = {10.1103/PhysRevLett.95.213001},
  url = {https://link.aps.org/doi/10.1103/PhysRevLett.95.213001}
}

@article{PhysRevLett.98.153003,
  title = {Strongly Correlated Two-Photon Transport in a One-Dimensional Waveguide Coupled to a Two-Level System},
  author = {Shen, Jung-Tsung and Fan, Shanhui},
  journal = {Phys. Rev. Lett.},
  volume = {98},
  issue = {15},
  pages = {153003},
  numpages = {4},
  year = {2007},
  month = {Apr},
  publisher = {American Physical Society},
  doi = {10.1103/PhysRevLett.98.153003},
  url = {https://link.aps.org/doi/10.1103/PhysRevLett.98.153003}
}

@article{PhysRevA.82.063816,
  title = {Waveguide QED: Many-body bound-state effects in coherent and Fock-state scattering from a two-level system},
  author = {Zheng, Huaixiu and Gauthier, Daniel J. and Baranger, Harold U.},
  journal = {Phys. Rev. A},
  volume = {82},
  issue = {6},
  pages = {063816},
  numpages = {10},
  year = {2010},
  month = {Dec},
  publisher = {American Physical Society},
  doi = {10.1103/PhysRevA.82.063816},
  url = {https://link.aps.org/doi/10.1103/PhysRevA.82.063816}
}

@article{Zeidan2026,
  title = {Superbunching from Coherently Driven Atoms in a Waveguide},
  author = {Zeidan, Zeidan and Karmstrand, Therese and Khanahmadi, Maryam and Johansson, G\"oran},
  journal = {Phys. Rev. Lett.},
  volume = {136},
  issue = {25},
  pages = {250803},
  numpages = {7},
  year = {2026},
  month = {Jun},
  publisher = {American Physical Society},
  doi = {10.1103/6fcg-2zns},
  url = {https://link.aps.org/doi/10.1103/6fcg-2zns}
}

@Article{Bhatti2015,
author={Bhatti, Daniel
and von Zanthier, Joachim
and Agarwal, Girish S.},
title={Superbunching and Nonclassicality as new Hallmarks of Superradiance},
journal={Scientific Reports},
year={2015},
month={Dec},
day={03},
volume={5},
number={1},
pages={17335},
doi={10.1038/srep17335},
url={https://doi.org/10.1038/srep17335}
}

@article{PhysRevA.95.033818,
  title = {Atom-light interactions in quasi-one-dimensional nanostructures: A Green's-function perspective},
  author = {Asenjo-Garcia, A. and Hood, J. D. and Chang, D. E. and Kimble, H. J.},
  journal = {Phys. Rev. A},
  volume = {95},
  issue = {3},
  pages = {033818},
  numpages = {16},
  year = {2017},
  month = {Mar},
  publisher = {American Physical Society},
  doi = {10.1103/PhysRevA.95.033818},
  url = {https://link.aps.org/doi/10.1103/PhysRevA.95.033818}
}

@article{PhysRevResearch.2.043213,
  title = {Atomic-waveguide quantum electrodynamics},
  author = {Masson, Stuart J. and Asenjo-Garcia, Ana},
  journal = {Phys. Rev. Res.},
  volume = {2},
  issue = {4},
  pages = {043213},
  numpages = {16},
  year = {2020},
  month = {Nov},
  publisher = {American Physical Society},
  doi = {10.1103/PhysRevResearch.2.043213},
  url = {https://link.aps.org/doi/10.1103/PhysRevResearch.2.043213}
}

@article{PhysRevResearch.3.033233,
  title = {Many-body localization in waveguide quantum electrodynamics},
  author = {Fayard, N. and Henriet, L. and Asenjo-Garcia, A. and Chang, D. E.},
  journal = {Phys. Rev. Res.},
  volume = {3},
  issue = {3},
  pages = {033233},
  numpages = {15},
  year = {2021},
  month = {Sep},
  publisher = {American Physical Society},
  doi = {10.1103/PhysRevResearch.3.033233},
  url = {https://link.aps.org/doi/10.1103/PhysRevResearch.3.033233}
}

@article{PhysRevX.10.031011,
  title = {Dynamics of Many-Body Photon Bound States in Chiral Waveguide QED},
  author = {Mahmoodian, Sahand and Calaj\'o, Giuseppe and Chang, Darrick E. and Hammerer, Klemens and S\o{}rensen, Anders S.},
  journal = {Phys. Rev. X},
  volume = {10},
  issue = {3},
  pages = {031011},
  numpages = {14},
  year = {2020},
  month = {Jul},
  publisher = {American Physical Society},
  doi = {10.1103/PhysRevX.10.031011},
  url = {https://link.aps.org/doi/10.1103/PhysRevX.10.031011}
}

@Article{Prasad2020,
author={Prasad, Adarsh S.
and Hinney, Jakob
and Mahmoodian, Sahand
and Hammerer, Klemens
and Rind, Samuel
and Schneeweiss, Philipp
and S{\o}rensen, Anders S.
and Volz, J{\"u}rgen
and Rauschenbeutel, Arno},
title={Correlating photons using the collective nonlinear response of atoms weakly coupled to an optical mode},
journal={Nature Photonics},
year={2020},
month={Dec},
day={01},
volume={14},
number={12},
pages={719-722},
issn={1749-4893},
doi={10.1038/s41566-020-0692-z},
url={https://doi.org/10.1038/s41566-020-0692-z}
}

@article{PhysRevA.107.013717,
  title = {Creation of nonclassical states of light in a chiral waveguide},
  author = {Kleinbeck, Kevin and Busche, Hannes and Stiesdal, Nina and Hofferberth, Sebastian and M\o{}lmer, Klaus and B\"uchler, Hans Peter},
  journal = {Phys. Rev. A},
  volume = {107},
  issue = {1},
  pages = {013717},
  numpages = {9},
  year = {2023},
  month = {Jan},
  publisher = {American Physical Society},
  doi = {10.1103/PhysRevA.107.013717},
  url = {https://link.aps.org/doi/10.1103/PhysRevA.107.013717}
}

@article{PhysRevA.97.023813,
  title = {Quantum interference and complex photon statistics in waveguide QED},
  author = {Zhang, Xin H. H. and Baranger, Harold U.},
  journal = {Phys. Rev. A},
  volume = {97},
  issue = {2},
  pages = {023813},
  numpages = {9},
  year = {2018},
  month = {Feb},
  publisher = {American Physical Society},
  doi = {10.1103/PhysRevA.97.023813},
  url = {https://link.aps.org/doi/10.1103/PhysRevA.97.023813}
}

@article{Zanner2022,
  author = {Zanner, Maximilian and Orell, Tuure and Schneider, Christian M. F. and Albert, Romain and Oleschko, Stefan and Juan, Mathieu L. and Silveri, Matti and Kirchmair, Gerhard},
  title = {Coherent control of a multi-qubit dark state in waveguide quantum electrodynamics},
  journal = {Nature Physics},
  year = {2022},
  volume = {18},
  number = {5},
  pages = {538--543},
  issn = {1745-2481},
  doi = {10.1038/s41567-022-01527-w},
  url = {https://doi.org/10.1038/s41567-022-01527-w}
}

@article{PhysRevA.82.043804,
  title = {Schemes for the observation of photon correlation functions in circuit QED with linear detectors},
  author = {da Silva, Marcus P. and Bozyigit, Deniz and Wallraff, Andreas and Blais, Alexandre},
  journal = {Phys. Rev. A},
  volume = {82},
  issue = {4},
  pages = {043804},
  numpages = {12},
  year = {2010},
  month = {Oct},
  publisher = {American Physical Society},
  doi = {10.1103/PhysRevA.82.043804},
  url = {https://link.aps.org/doi/10.1103/PhysRevA.82.043804}
}

@Article{Bozyigit2011,
author={Bozyigit, D.
and Lang, C.
and Steffen, L.
and Fink, J. M.
and Eichler, C.
and Baur, M.
and Bianchetti, R.
and Leek, P. J.
and Filipp, S.
and da Silva, M. P.
and Blais, A.
and Wallraff, A.},
title={Antibunching of microwave-frequency photons observed in correlation measurements using linear detectors},
journal={Nature Physics},
year={2011},
month={Feb},
day={01},
volume={7},
number={2},
pages={154-158},
doi={10.1038/nphys1845},
url={https://doi.org/10.1038/nphys1845}
}

@article{PhysRevLett.106.243601,
  title = {Observation of Resonant Photon Blockade at Microwave Frequencies Using Correlation Function Measurements},
  author = {Lang, C. and Bozyigit, D. and Eichler, C. and Steffen, L. and Fink, J. M. and Abdumalikov, A. A. and Baur, M. and Filipp, S. and da Silva, M. P. and Blais, A. and Wallraff, A.},
  journal = {Phys. Rev. Lett.},
  volume = {106},
  issue = {24},
  pages = {243601},
  numpages = {4},
  year = {2011},
  month = {Jun},
  publisher = {American Physical Society},
  doi = {10.1103/PhysRevLett.106.243601},
  url = {https://link.aps.org/doi/10.1103/PhysRevLett.106.243601}
}

@Article{Lu2021,
author={Lu, Yong
and Bengtsson, Andreas
and Burnett, Jonathan J.
and Wiegand, Emely
and Suri, Baladitya
and Krantz, Philip
and Roudsari, Anita Fadavi
and Kockum, Anton Frisk
and Gasparinetti, Simone
and Johansson, G{\"o}ran
and Delsing, Per},
title={Characterizing decoherence rates of a superconducting qubit by direct microwave scattering},
journal={npj Quantum Information},
year={2021},
month={Feb},
day={16},
volume={7},
number={1},
pages={35},
issn={2056-6387},
doi={10.1038/s41534-021-00367-5},
url={https://doi.org/10.1038/s41534-021-00367-5}
}

@Article{Kim2019,
author={Kim, May E.
and Chang, Tzu-Han
and Fields, Brian M.
and Chen, Cheng-An
and Hung, Chen-Lung},
title={Trapping single atoms on a nanophotonic circuit with configurable tweezer lattices},
journal={Nature Communications},
year={2019},
month={Apr},
day={09},
volume={10},
number={1},
pages={1647},
issn={2041-1723},
doi={10.1038/s41467-019-09635-7},
url={https://doi.org/10.1038/s41467-019-09635-7}
}

@article{
doi:10.1126/science.abi9917,
author = {Tamara Ðorđević  and Polnop Samutpraphoot  and Paloma L. Ocola  and Hannes Bernien  and Brandon Grinkemeyer  and Ivana Dimitrova  and Vladan Vuletić  and Mikhail D. Lukin },
title = {Entanglement transport and a nanophotonic interface for atoms in optical tweezers},
journal = {Science},
volume = {373},
number = {6562},
pages = {1511-1514},
year = {2021},
doi = {10.1126/science.abi9917},
URL = {https://www.science.org/doi/abs/10.1126/science.abi9917},
}

\end{document}